\documentclass[12pt]{article}\usepackage[]{graphicx}\usepackage[]{color}
\usepackage{alltt}
\usepackage{psfrag,epsf}
\usepackage{amsthm,amsmath,amssymb}
\usepackage{amsfonts}
\usepackage{multirow}
\usepackage{lscape}
\usepackage{setspace}
\usepackage{url}
\usepackage{enumerate}
\usepackage{caption}
\usepackage{subcaption}
\usepackage{authblk}
\usepackage{float}
\usepackage[toc,page]{appendix}

\allowdisplaybreaks[1]

\pdfoutput=1

\usepackage{layouts}

\usepackage{natbib}
\title{Dynamic Bayesian Influenza Forecasting in the United States with Hierarchical Discrepancy\footnote{LA-UR-17-22749}}
\author[1]{Dave Osthus}
\author[1]{James Gattiker}
\author[2]{Reid Priedhorsky}
\author[3]{Sara Y. Del Valle}
\affil[1]{Statistical Sciences, Los Alamos National Laboratory}
\affil[2]{High Performance Computing Environments, Los Alamos National Laboratory}
\affil[3]{Information Systems and Modeling, Los Alamos National Laboratory}
\small
\date{}
\IfFileExists{upquote.sty}{\usepackage{upquote}}{}
\begin{document}

\maketitle
\abstract{Timely and accurate forecasts of seasonal influenza would assist public health decision-makers in planning intervention strategies, efficiently allocating resources, and possibly saving lives. For these reasons, influenza forecasts are consequential. Producing timely and accurate influenza forecasts, however, have proven challenging due to noisy and limited data, an incomplete understanding of the disease transmission process, and the mismatch between the disease transmission process and the data-generating process. In this paper, we introduce a dynamic Bayesian (DB) flu forecasting model that exploits model discrepancy through a hierarchical model. The DB model allows forecasts of partially observed flu seasons to borrow discrepancy information from previously observed flu seasons. We compare the DB model to all models that competed in the CDC's 2015\textendash 2016 flu forecasting challenge. The DB model outperformed all models, indicating the DB model is a leading influenza forecasting model.}


\section{Introduction}
\label{sec:intro}

Influenza is a respiratory illness caused by the influenza virus that hospitalizes hundreds of thousands of people and affects millions in the United States annually \citep{fluburden2016}. Influenza also poses a significant burden on the U.S. economy through hospitalization costs and lost productivity from missing work \citep{molinari2007annual}. Flu surveillance is a collaborative effort between the Centers for Disease Control and Prevention (CDC) and many state and local healthcare providers, clinics and emergency rooms \citep{cdcsurveillance}. Monitoring the prevalence and geographic distribution of the flu is critical for targeted flu prevention strategies, such as vaccination campaigns and public education programs. 

In addition to flu monitoring, the CDC is also interested in flu forecasting. To better understand flu forecasting capabilities and to improve their usefulness to public health decision-makers, the CDC organized the first national flu forecasting competition in 2013 \citep{biggerstaff2016results}. Participation in the challenge included over a dozen models. The CDC defined forecasting targets relevant to public health decision-maker. These targets included the onset of the flu season, as well as the peak timing (PT) and peak intensity (PI) of the flu season. One-to-four week ahead forecasts (i.e., short term forecasts) were added as targets for the 2014\textendash 2015 challenge. From the 2013\textendash 2014 flu forecasting challenge, the CDC concluded that though flu forecasting is possible, much work remains. Flu forecasting is in its infancy and a concentrated effort to improve forecasting capabilities is needed in order for forecasts to be practically useful. The CDC has continued to organize an annual flu forecasting competition since the inaugural 2013\textendash 2014 challenge as a continuing effort to scope flu forecasting capabilities and provide an environment for collaboration and iterative improvement.

\cite{nsoesie2014systematic} and \cite{chretien2014influenza} provide reviews of the flu forecasting landscape. Flu forecasting models can be broadly categorized into four groups: mechanistic models, agent-based models, machine learning/regression models, and data-assimilation/dynamic models.
\begin{itemize}
\item \textbf{Mechanistic models} are differential-equation model descriptions of the disease transmission mechanism. They include a class of models referred to as compartmental models that partition a population into compartments and mathematically describe how individuals in the population move between compartments \citep[e.g.,][]{towers2009pandemic}.
\item \textbf{Agent-based models} simulate a population that mimics a real population using, for example, U.S. Census data to match various aspects of the simulated population to a real population (e.g., demographic information). The disease is then propogated via simulation through the simulated population and used to approximate the transmission of disease through a real population \citep[e.g.,][]{mniszewski2008episims,grefenstette2013fred}.
\item \textbf{Machine learning/regression models} are models that learn patterns in historical flu outbreaks and leverage those patterns for forecasting new flu seasons. This group includes such approaches as statistical time series \citep[e.g.,][]{soebiyanto2010modeling}, linear or regularized regression \citep[e.g.,][]{bardak2015prediction}, clustering \citep[e.g.,][]{viboud2003prediction}, and nonparametric approaches \citep[e.g.,][]{roni2015}. The machine learning/regression model approach to flu forecasting is characterized by the absence of a mechanistic model.
\item \textbf{Data-assimilation/dynamic models} usually involve embedding a mechanistic model into a probabilistic framework, allowing for the explicit modeling of the disease transmission process and observational noise \citep[e.g.,][]{osthus2017flu,hickmann2015forecasting,shaman2013,dukic2012tracking}. That is, the dynamic modeling approach combines two sources of uncertainty in the modeling; parametric uncertainty in the mechanistic model and random uncertainty in the observations.
\end{itemize}

Our modeling approach extends the data assimilation/dynamic modeling approach and can be viewed as a combination of the machine learning/regression approach and the data assimilation/dynamic modeling approach. Our model, referred to as a dynamic Bayesian (DB) model, explicitly accounts for systematic deviations between the mechanistic model and the data that are unable to be explained by pure observational noise. This systematic deviation is referred to as \emph{model discrepancy} and is modeled with a flexible, statistical model. Discrepancy modeling is an often used and effective modeling approach in the field of computer experiments, where systematic deviations between mechanistic models and data can be common \citep[e.g.,][]{kennedy2001bayesian,bayarri2007,higdon2008computer,brynjarsdottir2014learning}.

Including a discrepancy model is an appealing way to account for the systematic inadequacy of the mechanistic model. The basic insight leading to the inclusion of a discrepancy model in our DB model is that the disease transmission model and the data-generating model are not equivalent. Disease transmission is merely a component of the data-generating process. Thus, even if a mechanistic model were able to accurately identify the disease transmission process, there may still be a systematic discrepancy between the disease transmission model and the data, introducing a source of uncertainty unable to be accounted for by observational noise alone.

Though effective for data fitting, discrepancy modeling can make extrapolation (i.e., forecasting) challenging due to potential overfitting \citep{bayarri2007}. As opposed to previous Bayesian flu modeling approaches where flu tracking and parameter estimation were of interest \citep[e.g.,][]{coelho2011bayesian,dukic2012tracking}, our exclusive interest is forecasting. Thus, discrepancy modeling must be done with care. We address the discrepancy/forecasting issue by modeling the discrepancy hierarchically across all flu seasons. This allows us to borrow common discrepancy structure observed in past seasons in the forecasting of the flu for the current season. The hierarchical discrepancy model thus serves as a balance between the flexibility needed to match the statistical model to data and the structure needed to make useful and valid forecasts.

This paper makes contributions and advances in the following ways. 1) We introduce and demonstrate the importance of discrepancy modeling to the growing and consequential field of flu forecasting. Discrepancy modeling is done hierarchically, allowing information to be shared across available flu seasons. 2) We demonstrate the superiority of our approach relative to all models that competed in the CDC's 2015\textendash 2016 flu forecasting competition, providing yet another instance where discrepancy modeling is not only conceptually appealing but also practically effective. 3) In an effort to advance flu forecasting capabilities, much effort has been spent identifying possibly useful, nontraditional data sources such as Google \citep{ginsberg2009detecting} and Wikipedia \citep{generous2014global}. Alternatively, as we demonstrate, flu forecasting can be improved through carefully made modeling choices, making use of the the available traditional data hierarchically. 4) The DB model is implemented in standard Bayesian software, JAGS, allowing others to critique, extend, and improve upon the model.


The paper is laid out as follows. In Section \ref{sec:data}, we present the data. In Sections \ref{sec:sir} and \ref{sec:model}, we present the mechanistic model and statistical DB model, respectively. We investigate and assess the DB model in Sections \ref{subsec:fitex} and \ref{subsec:assessment}, respectively. The DB model is compared to all participating models in the 2015\textendash 2016 flu forecasting challenge organized by the CDC in Section \ref{subsec:comparison}. We conclude with a discussion in Section \ref{sec:discussion}.

\section{Data}
\label{sec:data}
The CDC performs influenza surveillance in the United States via a multitude of surveillance efforts including virologic, outpatient, mortality, and hospitalization surveillance systems \citep{cdcsurveillance}. In this paper, we focus exclusively on outpatient illness surveillance. Symptomatic information on patient visits to healthcare providers is collected through the United States outpatient influenza-like illness surveillance network (ILINet). ILINet is a collection of almost 3,000 healthcare providers across the United States. These participating healthcare providers supply information to the CDC regarding the number of patients seen for any reason and the number of patients seen with a diagnosed influenza-like illness on a weekly basis. An influenza-like illness is defined as a temperature greater than or equal to 100 degrees Fahrenheit and a cough or sore throat with no known cause other than influenza. It is important to note an influenza-like illness diagnosis and an influenza diagnosis are not equivalent. Many diseases have flu-like symptoms prior to fully developing such as measles, rubella, tuberculosis, food poisoning, dengue, and malaria. ILINet is ill-equipped and not designed to discriminate between the flu and diseases with flu-like symptoms. A flu diagnosis requires some form of laboratory test. We model and forecast influenza-like illness in this paper.

The CDC aggregates, organizes, and ultimately releases influenza-like illness information to the public weekly throughout the year at both the national and health and human service region levels \citep{hhs}. In this paper, we focus exclusively on national level influenza-like illness surveillance. The proportion of the population with an influenza-like illness is estimated by the CDC with the quantity weighted influenza-like illness (wILI) where wILI is, ``the percentage of patient visits to healthcare providers for influenza-like illness reported each week weighted on the basis of state population," \citep{cdcsurveillance}.

\begin{figure}[!ht]
\centering
\includegraphics[width=.7\textwidth]{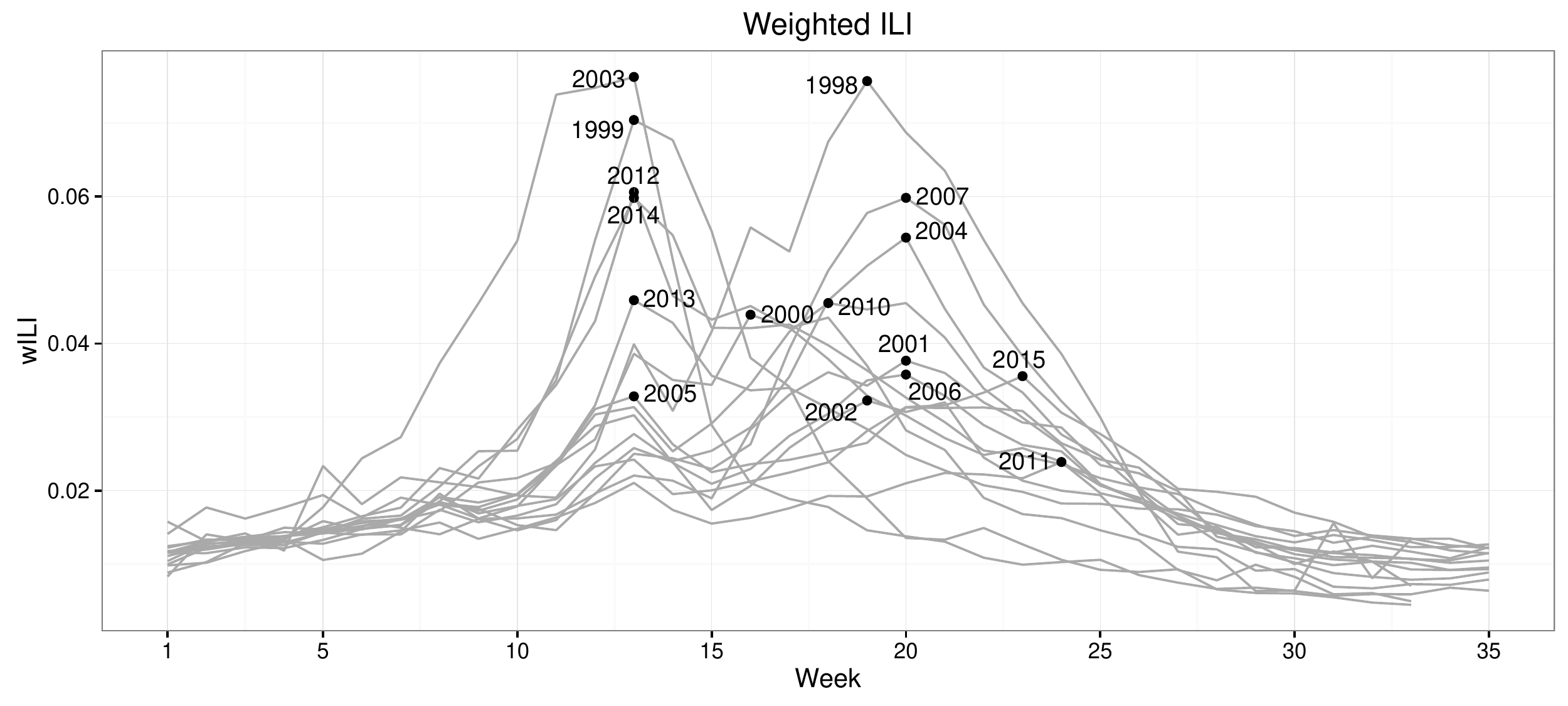}
\caption{Weighted ILI for flu seasons 1998 through 2015, sans H1N1 seasons 2008 and 2009. Grey lines correspond to flu season trajectories. Week 1 is roughly the first week of October, while week 35 is roughly the last week of May. The black points are located at the peak timing (x-axis) and peak intensity (y-axis) of their respective flu season. Peak timing occurs between week 13 (roughly the end of December) and week 24 (roughly the middle of March).}
\label{fig:wili}
\end{figure}

Figure \ref{fig:wili} shows wILI for seasonal influenza from 1998 through 2015, excluding the pandemic H1N1 flu seasons 2008 and 2009. We focus on seasonal influenza rather than the more severe and substantially less frequent pandemic flu seasons as seasonal and pandemic flu transmission dynamics are appreciably different. Seasonal flu outbreaks have a relatively predictable profile, as can be seen in Figure \ref{fig:wili}. The population often has partial immunity to the circulating virus(es) of seasonal influenza and it occurs nearly ever year in the United States. Pandemic influenza follows much less predictable transmission patterns, due in part to the relatively low immunity in the population to the new, yet to be seen strain of influenza. As a result, pandemic influenza is typically non-recurring. There have been four instances of pandemic influenza since 1900: 1918, 1957, 1968, and 2009 \citep{cdcpandemic}.

A typical flu season begins in October and lasts until as late as May, thus, the 1998 flu season refers to the season starting in 1998 and ending in 1999. In this paper, flu season week 1, referred to as ``week 1," corresponds to Morbidity and Mortality Weekly Report (MMWR) week 40. MMWR is a common epidemiological dating system used for reporting purposes \citep{cdcmmwr}. Week 1 roughly corresponds to the first week of October while week 35 roughly corresponds to the last week of May.

\begin{figure}[!ht]
\centering
\includegraphics[width=.9\textwidth]{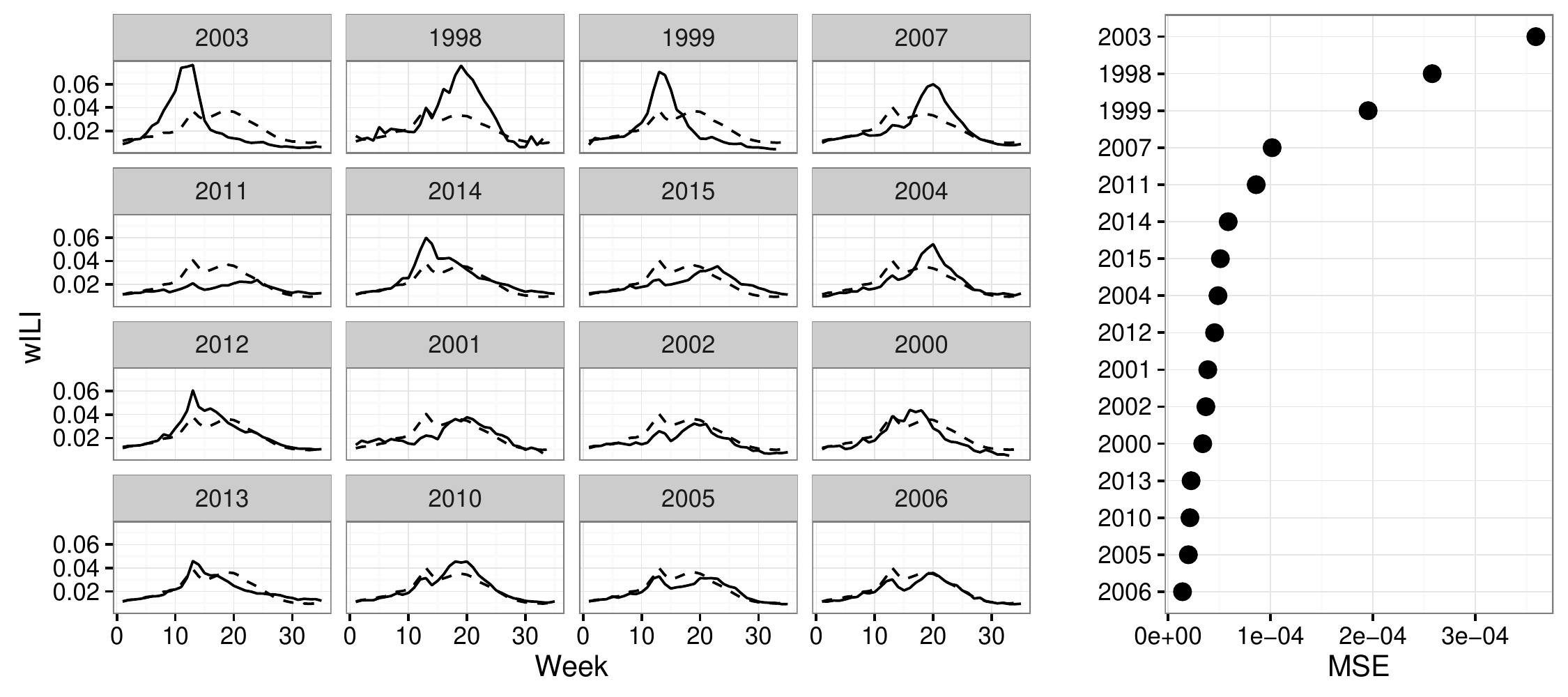}
\caption{(Left) Weighted ILI (solid line) for each season and the average wILI for all other flu seasons (dashed line). Seasons are ordered from top, left to bottom, right by decreasing mean-squared error (MSE). (Right) The MSE between each flu season and the average of the other flu seasons. Flu seasons 2003, 1998, and 1999 are the most ``atypical" flu seasons.}
\label{fig:msewili}
\end{figure}

In Figure \ref{fig:wili}, the point corresponds to the peak timing (PT) and peak intensity (PI) of each flu season. We see most flu seasons either peak early (six flu seasons peaked on week 13 -- roughly the end of December) or late (nine flu season peaked between week 18 and 24 -- roughly the beginning of February through the middle of March), with the 2000 flu season peaking during week 16. All flu seasons in Figure \ref{fig:wili} exhibit a similar pattern; wILI is low at the beginning of the flu season, increases to a maximum in the middle of the flu season, and reverts to low levels by the end of the flu season. Though each season shares this general pattern, heterogeneity exists between flu seasons. Some flu seasons appear to deviate from ``typical" flu seasons more than others. This observation is illustrated in Figure \ref{fig:msewili}. Flu seasons 1998, 1999, and 2003 most significantly deviate from ``typical" as calculated by the mean-squared error (MSE) between each flu season and the week-specific average of all other flu seasons. Other flu seasons, such as 2005, 2006, and 2010 are the most ``typical" by this same measure.

\section{Susceptible-Infectious-Recovered Model}
\label{sec:sir}

\begin{figure}[!ht]
\centering
\includegraphics[width=.9\textwidth]{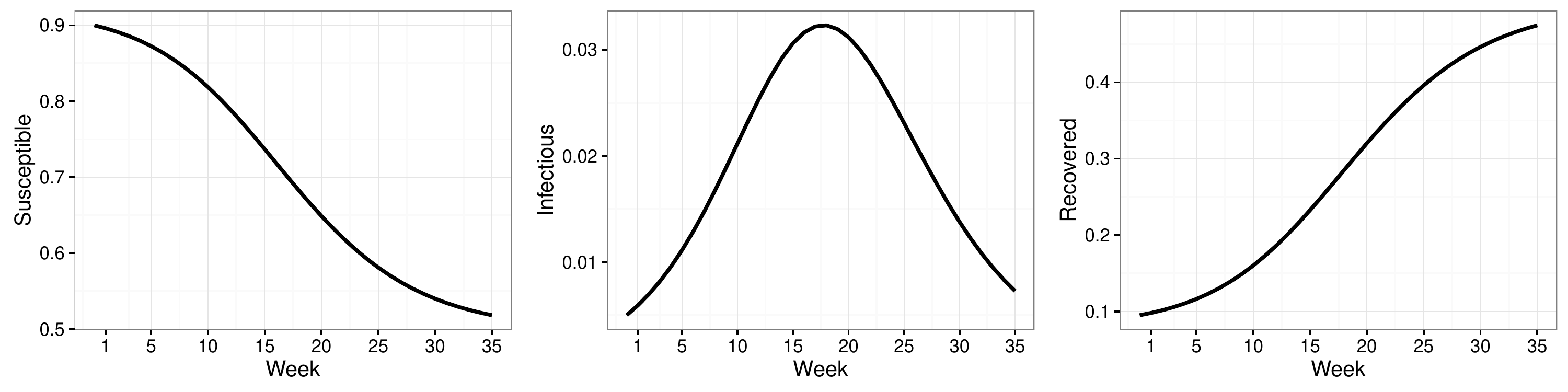}
\caption{Solution to an SIR model with $S_0 = 0.9$, $I_0 = 0.005$, $R_0 = 0.095$, $\gamma = 0.55$, and $\beta = 0.8$, where $S_0$, $I_0$, and $R_0$ are the proportions of the population susceptible, infectious, and recovered from the disease at time 0.}
\label{fig:sirex}
\end{figure}

The susceptible-infectious-recovered (SIR) model was introduced in the seminal work of \cite{kermack1927contribution} and is considered the foundation for modern public health \citep{weiss2013sir}. The SIR model is a mechanistic model that describes how an infectious disease spreads through a closed population via the following set of nonlinear, ordinary differential equations:
\begin{align}
\label{eq:sir}
\frac{dS}{dt} &= -\beta S I, &
\frac{dI}{dt} &= \beta S I - \gamma I, &
\frac{dR}{dt} &= \gamma I,
\end{align}

\noindent where $\beta > 0$ is the disease transmission rate and $\gamma > 0$ is the recovery rate. Under the SIR model, individuals in the population are partitioned into three mutually exclusive and exhaustive compartments: a susceptible, infectious, and recovered compartment. The SIR model describes the rate at which individuals move from being susceptible to infectious and finally recovered. $S$, $I$, and $R$ in Equation \ref{eq:sir} represent proportions of the population, such that $S + I + R = 1$ for all times. An example of an SIR trajectory is shown in Figure \ref{fig:sirex}.

The trajectories of the susceptible and recovered SIR compartments are monotonically non-increasing and non-decreasing, respectively. All SIR infectious trajectories can be partitioned into two designations: epidemics and non-epidemics. These designations are completely determined by the relationship between $S_0$, the proportion of the population initially susceptible to the disease, and $\rho$, where $\rho = \gamma/\beta$ \citep{weiss2013sir}. An epidemic designation occurs when $S_0 > \rho$. An epidemic graphically corresponds to an infectious trajectory that monotonically increases to a maximum followed by a monotonic decrease to zero as time goes to infinity. The infectious trajectory shown in Figure \ref{fig:sirex} is an epidemic. A non-epidemic designation occurs when $S_0 \leq \rho$, graphically meaning the infectious trajectory monotonically decreases from $I_0$, the proportion of the population initially infectious with the disease, to zero as time goes to infinity. As can be seen in Figure \ref{fig:wili}, every flu season we consider exhibits the general shape of an epidemic, where wILI is low at week 1 of the flu season, increases to a maximum between weeks 13 and 24, and declines to low levels by week 35.

\begin{figure}[!ht]
\centering
\includegraphics[width=.9\textwidth]{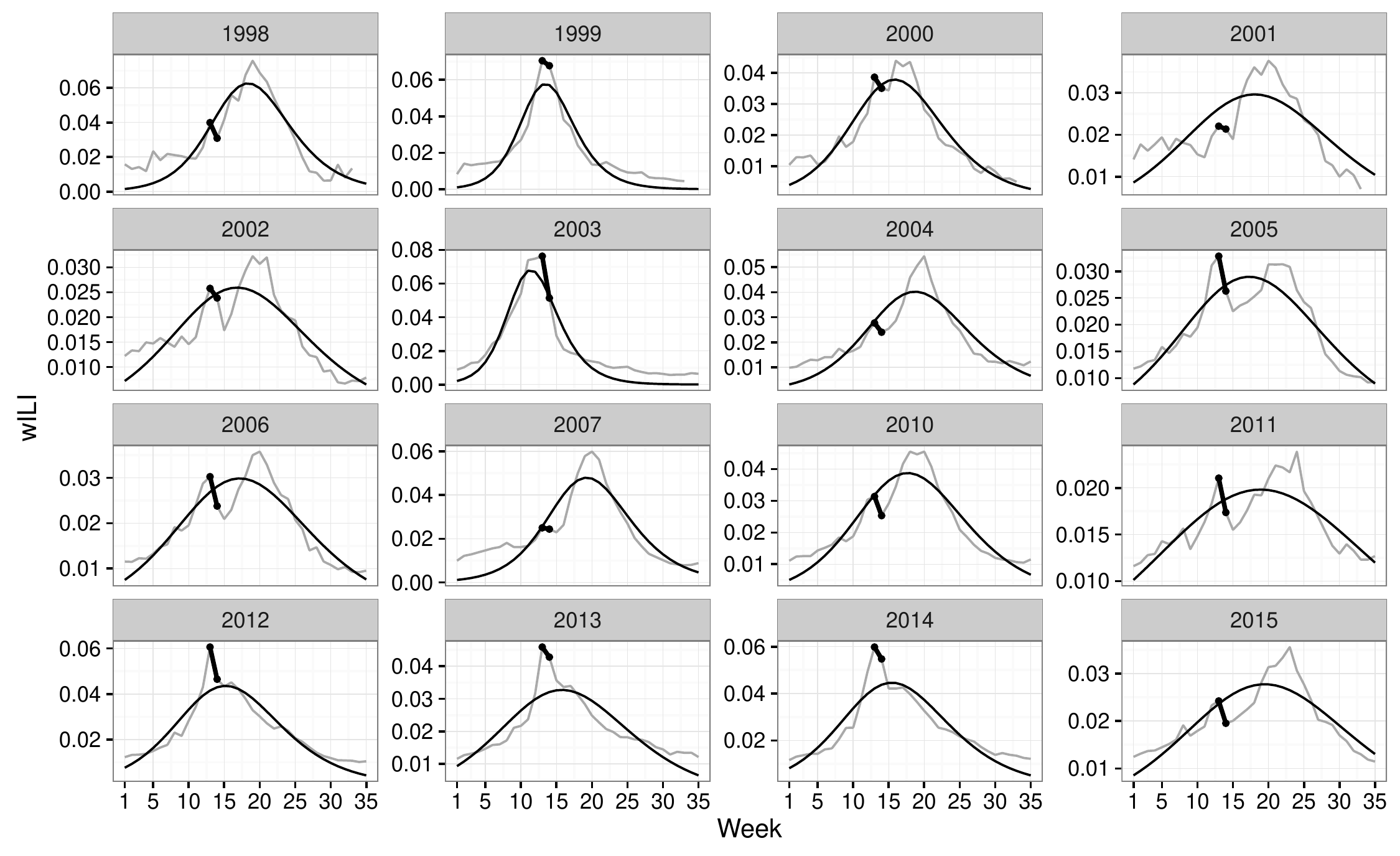}
\caption{Fitted SIR model (black curve) to wILI (grey) by flu season. The black line segment denotes the systematic decline in wILI from week 13 to week 14 in all flu seasons.}
\label{fig:sirfit}
\end{figure}

Figure \ref{fig:sirfit} plots the best fit SIR trajectory and wILI for each flu season constrained to $S_0=0.9$. We see the SIR trajectories match the general shape of wILI. These trajectories do not, however, replicate some of the more nuanced structure of wILI. For instance, some seasons exhibit a double peak (e.g., 2002, 2005, 2006, and 2011). The SIR model, however, is incapable of capturing two peaks within a flu season; it can only capture one. Another interesting feature of wILI happens at weeks 13 and 14. In every flu season, wILI on week 13 is larger than wILI on week 14. For some flu seasons, this downturn in wILI from week 13 to 14 signifies the peak of the flu season; flu seasons 1999, 2003, 2005, 2012, 2013, 2014 all peak on week 13. For the other flu seasons, however, the decline in wILI from week 13 to week 14 does not signify the peak of the flu season as they all exhibit peaks later in the season. It is not known exactly why wILI reliably declines from week 13 to 14, though explanations have been posited, such as a change in disease transmission during winter holidays \citep{ewing2016contact,garza2013effect,huang2014us}. For the purposes of forecasting wILI, what is important is that the decline in wILI from week 13 to 14 is reliable and unable to be captured by the SIR model.

\begin{figure}[!ht]
\centering
\includegraphics[width=.7\textwidth]{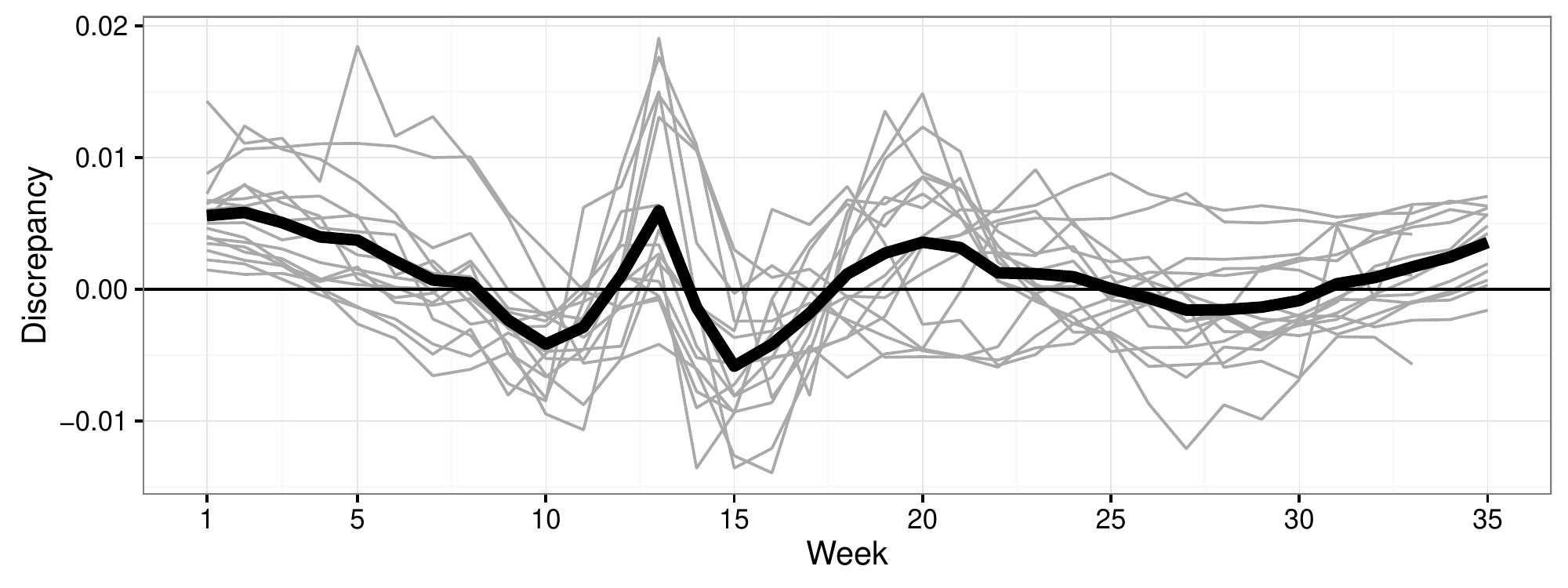}
\caption{Discrepancy trajectories for all flu seasons are denoted by grey lines. The black line is the average discrepancy trajectory. Discrepancy tends to be greater than zero and the beginning (week 1), peak (weeks 13 and 20), and end of the season (week 35).}
\label{fig:sirresids}
\end{figure}

Figure \ref{fig:sirresids} plots the residuals between wILI and the fitted SIR curves for all flu seasons in Figure \ref{fig:sirfit}, referred to as discrepancy trajectories, along with the average discrepancy trajectory. Figure \ref{fig:sirresids} articulates the systematic deviations between best fit SIR trajectories and wILI. The SIR model tends to under estimate wILI near weeks at the start (week 1), end (week 35), and peak (weeks 13 and 20) of the flu season, while systematically over estimating wILI near weeks 10, 15, and 27.

Figure \ref{fig:sirfit} suggests describing wILI with even the best fitting SIR model is inadequate. Furthermore, Figure \ref{fig:sirresids} suggests the discrepancy between the best fit SIR model and wILI cannot plausibly be described by random error alone as there is structure across both seasons and time in the discrepancy. In the next section, we present the dynamic Bayesian model which incorporates both the SIR model and the structured discrepancy hierarchically for the ultimate purpose of forecasting future wILI.

\section{Dynamic Bayesian Model}
\label{sec:model}
In this section, we describe the various components of the DB model, broadly partitioned into the data model (Section \ref{subsec:datamodel}) and the process model (Section \ref{subsec:processmodel}).

\subsection{Data Model}
\label{subsec:datamodel}
Let $y_{j,t}$ be wILI for flu season $j = 1998, \ldots, 2007, 2010, \ldots, 2015$ during week $t=1, 2, \ldots, T$ where $T=35$. We model the proportion $y_{j,t}$ as
\begin{align}
\label{eq:datamodel}
y_{j,t} &\sim \text{Beta}(\lambda \pi_{j,t}, \lambda (1-\pi_{j,t})),
\end{align}

\noindent where $\pi_{j,t} \in [0,1]$ for all $j$ and $t$ is the true but unobservable proportion of the population with an influenza-like illness. The mean and standard deviation for $y_{j,t}$ are
\begin{align}
\label{eq:datamoments}
\text{E}(y_{j,t}) &= \pi_{j,t},\\
\text{SD}(y_{j,t}) &= \Bigg(\frac{\pi_{j,t}(1-\pi_{j,t})}{1+\lambda}\Bigg)^{0.5}.
\end{align}


\noindent Equation \ref{eq:datamodel} assumes $y_{j,t}$ is unbiased for $\pi_{j,t}$. The concentration parameter $\lambda$ governs the standard deviation of $y_{j,t}$. That is, $\lambda$ governs the random variability of $y_{j,t}$ caused by such things as sampling variability, ILI diagnosis errors, and reporting variability. For a given $\lambda$, the standard deviation increases with increasing $\pi_{j,t}$. Increasing random variability with increasing levels of $\pi_{j,t}$ is a desired feature other models have attempted to mimic through ad hoc means \cite[e.g.,][]{shaman2012}. The Beta distribution is able to capture this feature naturally. The random variability is not expected to vary across flu seasons and is poorly learned from the data. For these reasons, we set $\lambda = 4,500$. This choice of $\lambda$ implies the standard deviation of $y_{j,t}$ is 0.0025 when $\pi_{j,t}$ is equal to 0.03 and $y_{j,t}$ is 0.0035 when $\pi_{j,t}$ is equal to 0.06.


\subsection{Process Model}
\label{subsec:processmodel}

We model the logit of the true but unobservable proportion of influenza-like illness, $\pi_{j,t}$, as the sum of three components,
\begin{align}
\label{eq:processmodel}
\text{logit}(\pi_{j,t}) &= \text{logit}(I_{j,t}) + \mu_t + \delta_{j,t}.
\end{align}

\noindent Equation \ref{eq:processmodel} decomposes logit$(\pi_{j,t})$ into an SIR model component $\text{logit}(I_{j,t})$ a discrepancy component common to all flu seasons $\mu_t$ and a discrepancy component specific to each flu season $\delta_{j,t}$. The SIR model component represents the component of $\pi_{j,t}$ that can be described by the SIR model. Ideally, logit($I_{j,t}$) would describe all of logit($\pi_{j,t}$) implying the discrepancy terms are zero. On the basis of Figure \ref{fig:sirfit}, though, we know the SIR model cannot capture all the relevant features of $y_{j,t}$ and by extension, $\pi_{j,t}$. Thus, the common discrepancy component $\mu_t$ captures systematic differences between logit$(\pi_{j,t})$ and logit$(I_{j,t})$ shared by all flu seasons. We anticipate there is discrepancy structure common to all flu seasons on the basis of the non-zero, average discrepancy trajectory in Figure \ref{fig:sirresids}. The flu season-specific discrepancy term, $\delta_{j,t}$, captures the component of logit($\pi_{j,t}$) unexplained by logit($I_{j,t}$) and $\mu_t$. Again, we anticipate season-specific discrepancy is needed on the basis of the season-specific discrepancy trajectories in Figure \ref{fig:sirresids}. In what follows, we specify the statistical models for each of the components of Equation \ref{eq:processmodel}.

\subsubsection{Model for logit($I_{j,t}$)}
\label{subsubsec:infectious}

We model $I_{j,t}$, the infectious proportion of the population according to the SIR model for flu season $j$ during week $t$, as the solution to Equation \ref{eq:sir}. An explicit formula solution to Equation \ref{eq:sir}, however, is unavailable. Thus, a numerical approximation method is used. We follow \cite{osthus2017flu} and use the fourth order Runge-Kutta approximation method (RK4) to approximate the solution to the SIR model. The details of the RK4 method can be found in Appendix \ref{appendix:rk4}. The RK4 approximation method is known to be more stable than the simpler Euler's method; a result we have also found to be true.

\begin{figure}[!ht]
\centering
\includegraphics[width=.9\textwidth]{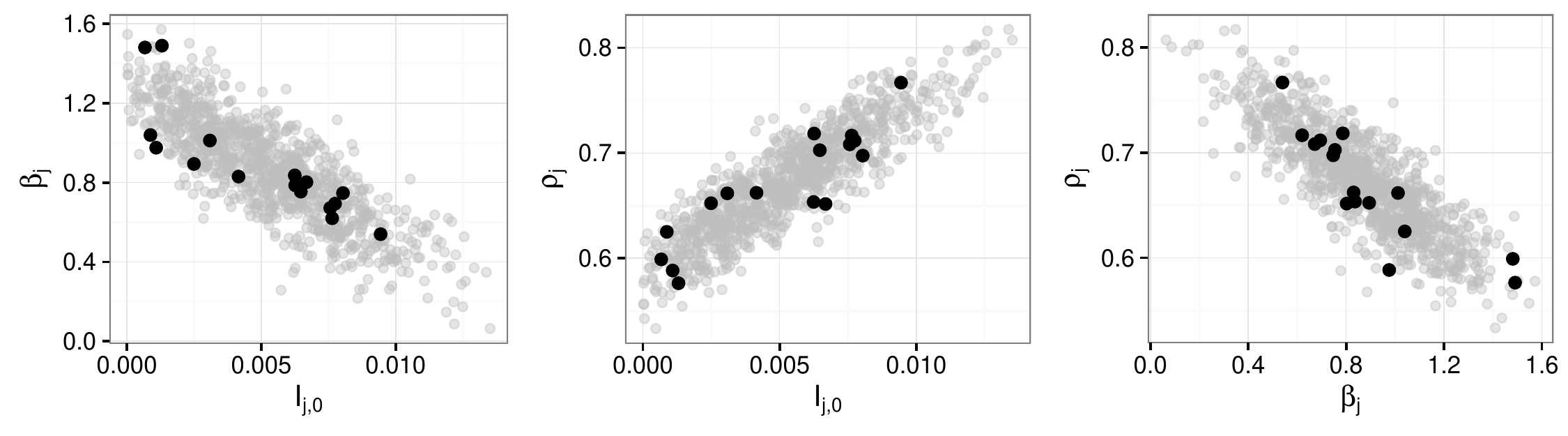}
\caption{Parameter estimates for the best fit SIR models corresponding to each flu season in Figure \ref{fig:sirfit} (black) and 1,000 draws from the truncated Gaussian prior for $I_{j,0}$, $\beta_j$, and $\rho_j$ (grey).}
\label{fig:sirparams}
\end{figure}

Examination of Equations \ref{eq:rk4} and \ref{eq:k} in Appendix \ref{appendix:rk4} makes it clear that $I_{j,t}$ is completely determined once $S_{j,0}$, $I_{j,0}$, $R_{j,0}$, $\gamma_j$, and $\beta_j$ are specified. We set $S_{j,0}=0.9$ for all $j$ following \cite{osthus2017flu} as there is little information to learn about the susceptible and recovered trajectories of the SIR model from only wILI \citep{capaldi2012}. Setting $S_{j,0}=0.9$ is, thus, an identifiability constraint. We assign an informative prior to $I_{j,0}$, $\beta_j$ and $\rho_j$ via empirical Bayes by fitting a multivariate Gaussian distribution to the parameter estimates of the fitted SIR models shown in Figure \ref{fig:sirfit} and truncating to respect known and/or assumed boundary constraints. SIR parameter estimates and draws from the prior are shown in Figure \ref{fig:sirparams}. The truncation for $I_{j,0}$ is $(0,0.1)$ ensuring that $I_{j,0}$ is positive and $S_{j,0} + I_{j,0} \leq 1$. To maintain mass balance, we set $R_{j,0}$ equal to $1-S_{j,0}-I_{j,0}$. The truncation for $\beta_j$ is $(0,\infty)$ ensuring it is positive. The truncation for $\rho_j$ is $(0,.9)$, ensuring it is positive and less than $S_{j,0}$, thus restricting the SIR model to be an epidemic. We find the upper bound truncation on $\rho_j$ helps with numerical stability in our posterior sampling in addition to aligning with our expectations for the shape of the SIR infectious trajectory.

We emphasize that the ultimate goal of the model is to predict future observations of wILI for a completely or partially unobserved flu season where data are not yet available. Data from partially observed flu seasons are not used in the prior specification of $I_{j,0}$, $\beta_j$, and $\rho_j$.  For example, if the model is forecasting wILI for flu season 2015, then SIR parameter estimates from all non-2015 flu seasons are used to estimate the parameters for the prior on $I_{j,0}$, $\beta_j$, and $\rho_j$.

The prior for $I_{j,0}$, $\beta_j$, and $\rho_j$ assumes that the best fitting SIR model for a completely or partially unobserved flu season comes from the same distribution as the best fitting SIR models for completely observed seasons. That is, we have an informative prior about the initial conditions and parameters of the best fitting SIR model prior to observing any data. We believe this is a reasonable assumption for the task of forecasting seasonal influenza. This, however, would be a questionable assumption for the task of forecasting a less predictable and non-recurrent disease, such as pandemic influenza.

\subsubsection{Model for $\mu_t$}
\label{subsubsec:mu}
The discrepancy process $\mu_t$ captures the systematic discrepancy common to all flu seasons and is what allows the forecasts of a partially observed season to borrow discrepancy information from other flu seasons. We specify the $\mu_t$ process as a reverse-random walk. Specifically,
\begin{align}
\label{eq:muprocessT}
\mu_T &\sim \text{N}(0,\sigma^2_{\mu_T}),\\
\label{eq:muprocesst}
\mu_t|\mu_{t+1} &\sim \text{N}(\mu_{t+1},\sigma^2_{\mu}).
\end{align}

\noindent The random-walk specification is a way to impose temporal structure to the common discrepancy model as $\mu_t$ depends on $\mu_{t+1}$. The \emph{reverse} random-walk specification is related to how wILI is released. Within a flu season, wILI is first available for week 1, then week 2, and so forth. As a result, we are always forecasting the end of the flu season and seldom the beginning of the flu season. From Figure \ref{fig:wili}, it is clear that the end of the flu season is relatively well-behaved and predictable as compared to the middle of the flu season. That is, though there is considerable uncertainty regarding the trajectory wILI will take, there is much less uncertainty regarding wILI's destination on week 35. A reverse-random walk helps bridge the gap between the last wILI observation and the end of the flu season. A reverse random-walk has been used with success in other forecasting contexts, such as presidential election forecasting \citep{linzer2013}.

We assigned the following priors to the precisions of Equations \ref{eq:muprocessT} and \ref{eq:muprocesst}:
\begin{align}
\label{eq:muprecisionT}
\sigma^{-2}_{\mu_T} &\sim \text{Gamma}(2,2), &
\sigma^{-2}_{\mu} &\sim \text{Gamma}(2,0.02).
\end{align}
\noindent The priors reflect a belief that $\sigma_{\mu_T}$ will be larger than $\sigma_{\mu}$.

\subsubsection{Model for $\delta_{j,t}$}
\label{subsubsec:delta}

We would like to explain logit$(\pi_{j,t})$ with logit$(I_{j,t})$ and $\mu_t$ if possible. Figure \ref{fig:sirresids}, however, suggests this is not possible and thus the model will likely benefit from a season-specific discrepancy term, $\delta_{j,t}$. The introduction of $\delta_{j,t}$ creates an issue with model identifiability. As an identifying constraint, we set
\begin{align}
\label{eq:deltaprocessinit}
\delta_{j,T} &= -\text{logit}(I_{j,T}).
\end{align}

\noindent The constraint in Equation \ref{eq:deltaprocessinit} implies,
\begin{align}
\label{eq:yimplication}
y_{j,T} &\sim \text{Beta}(\lambda \text{logit}^{-1}(\mu_T), \lambda(1-\text{logit}^{-1}(\mu_T))),
\end{align}
\noindent as $\pi_{j,T} = \text{logit}^{-1}(\mu_T)$. That is, the data model for wILI on week $T$ is a function of $\mu_T$, the discrepancy component common to all flu seasons, and $\lambda$, the parameter governing the random variability of the data model.

We impose temporal structure on $\delta_{j,t}$ and encourage it to be close to zero for all $t \neq T$ by modeling $\delta_{j,t}$ with the following autoregressive, reverse random-walk:
\begin{align}
\label{eq:deltaprocess}
\delta_{j,t}|\delta_{j,t+1} &\sim \text{N}(\alpha_j \delta_{j,t+1},\sigma^2_{\delta,j}).
\end{align}
\noindent The autoregressive parameter $\alpha_j \in [0,1]$ encourages $\delta_{j,t}$ to be close to zero. As with $\mu_t$, the random-walk structure for $\delta_{j,t}$ imposes a temporal dependence, as $\delta_{j,t}$ depends on $\delta_{j,t+1}$. The \emph{reverse} random-walk allows for easy incorporation of the identifying constraint in Equation \ref{eq:deltaprocessinit}.

Both $\alpha_j$ and the variance $\sigma^2_{\delta,j}$ are modeled hierarchically. The model for $\alpha_j$ is,
\begin{align}
\label{eq:alpha}
\text{logit}(\alpha_j) &\sim \text{TN}_{(\text{logit}(0.02),\text{logit}(0.98))}(\text{logit}(0.9),\sigma^2_{\alpha}),\\
\sigma_{\alpha} &\sim \text{Gamma}(2,2),
\end{align}
\noindent where TN$_{(\text{logit}(0.02),\text{logit}(0.98))}$ is a truncated Gaussian distribution, truncated between logit($0.02$) and logit($0.98$). The truncation improved numerical stability of the posterior sampler. The mean in Equation \ref{eq:alpha} reflects our prior belief that $\alpha_j$ is near 0.9 for all $j$. Note that if $\alpha_j = 0$, then Equation \ref{eq:deltaprocess} has no random-walk structure, as each $\delta_{j,t}$ is modeled as a mean 0 Gaussian distribution with a season-specific variance. Said another way, the closer $\alpha_j$ gets to zero, the less temporal structure exists in the season-specific discrepancy process.

Finally, we assign the following hierarchical prior to the precisions of the season-specific discrepancy model, $\sigma^{-2}_{\delta,j}$:
\begin{align}
\label{eq:deltavar}
\sigma^{-2}_{\delta,j} &\sim \text{Gamma}(a_{\delta},b_{\delta}),
\end{align}
\noindent where $a_{\delta} \sim \text{Gamma}(5,1)$ and $b_{\delta} \sim \text{Gamma}(1,10)$. The parameter $\sigma^{-2}_{\delta,j}$ is a flexibility parameter. The smaller $\sigma^{-2}_{\delta,j}$ becomes (or equivalently, the larger the variance $\sigma^2_{\delta,j}$ becomes), the more flexible $\delta_{j,t}$ becomes. Thus, we expect flu seasons that more acutely deviate from ``typical" wILI behavior to require larger variances, $\sigma^2_{\delta,j}$, than more ``typical" flu seasons. The hierarchical specification for precisions $\sigma^{-2}_{\delta,j}$ allows the estimation of $\sigma^{-2}_{\delta,j}$ for a partially observed flu season to borrow information from fully observed flu seasons.

\section{Results}
\label{sec:results}
In practice, the DB model is updated each week when new wILI data becomes available. We take a ``leave-one-season-out" approach to forecasting, where we make use of all available data from the seasons \emph{not} being forecasted as well as all of the \emph{observed} data from the season being forecast. This approach mimics how forecasting is done in real-time. We refer to each model fit with a ``Season.Week" naming convention. Model ``Season.Week" refers to a model fit to all observations from flu seasons not equal to ``Season" and all observations in ``Season" from week 1 through week ``Week". For example, model 2015.3 is a model fit to all wILI observations not in flu season 2015 plus weeks 1 through 3 of 2015 (recalling flu season 2015 means the 2015\textendash 2016 flu season). Forecasting model 2015.3 means forecasting the unobserved data for weeks 4 through 35 of flu season 2015.

The posterior sampling of the DB model was performed using the \verb|rjags| package \citep{rjags} within the \textsf{R} programming language \citep{R}, which calls the software ``Just Another Gibbs Sampler," or JAGS \citep{plummer2003}.

Model parameter convergence was checked for the 2015.3 model by running four chains for 100,000 iterations, throwing away the first half as burn-in and thinning every 20th iteration, resulting in four chains each of length 2,500. We assessed MCMC convergence for all latent quantities of the model with the Gelman-Rubin diagnostic, $\hat{R}$ \citep{gelman1992}. $\hat{R}$ was computed using the \verb|gelman.diag()| function in the \verb|coda| package \citep{coda}. All $\hat{R}$s were less than 1.1, suggesting no evidence for lack of convergence. For all other models, we ran one chain for 50,000 iterations, throwing away the first half as burn-in and thinning every 10th iteration, resulting in a chain of 2,500.

In Section \ref{subsec:fitex}, we illustrate how the model is updated each week for the 2015 flu season and discuss the different model components. We assess forecasts for all seasons in the context of predictive empirical coverage in Section \ref{subsec:assessment}. Finally, in Section \ref{subsec:comparison}, we compare the DB model's forecasting accuracy to the 14 flu forecasting models that participated in the CDC's 2015\textendash 2016 flu forecasting challenge.

\subsection{Model Fit to 2015\textendash 2016 Flu Season}
\label{subsec:fitex}

\begin{figure}[!ht]
\centering
\includegraphics[width=.7\textwidth]{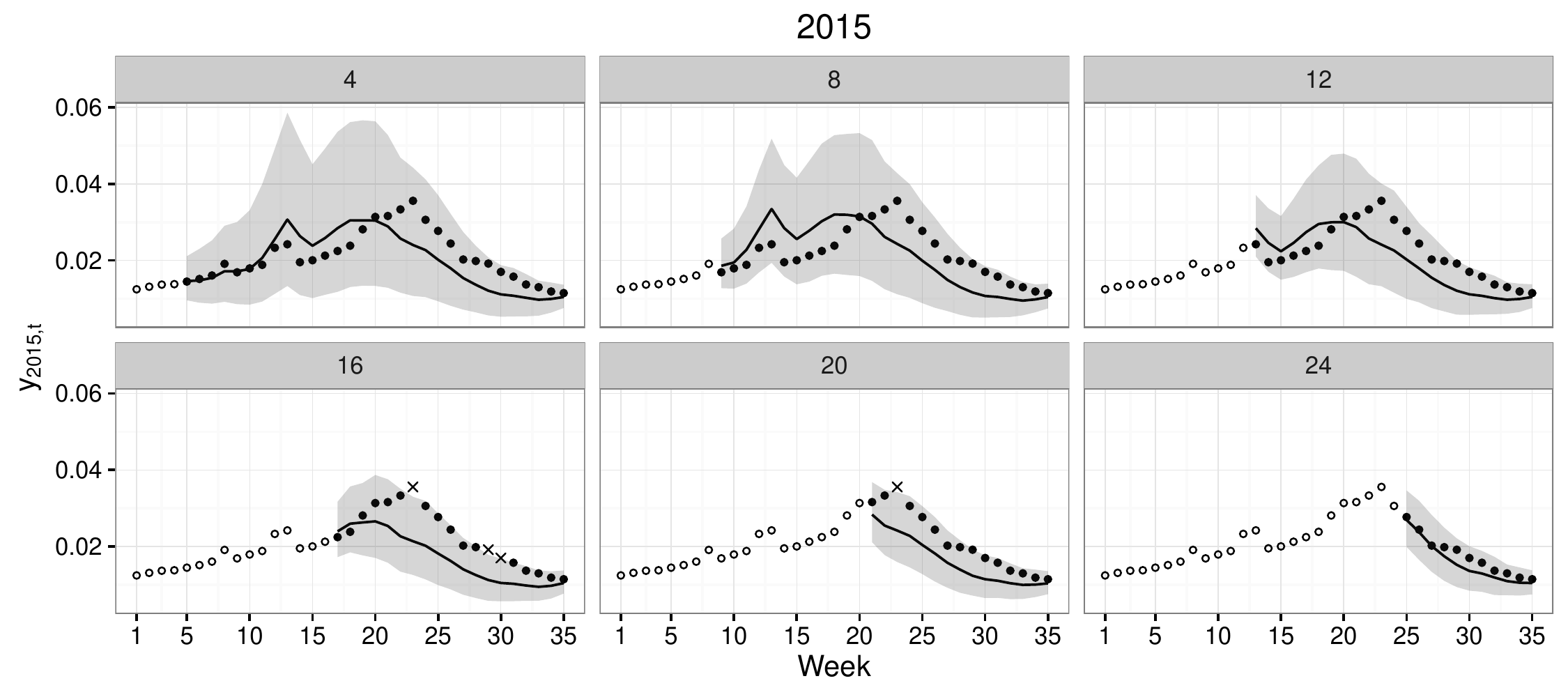}
\caption{Forecasts for the 2015 flu season. The panel label number denotes the ``Week" of the ``2015.Week" model fit. Black, hollow circles denote observed data. Grey bands denote 95\% point-wise posterior predictive intervals for unobserved data while the black line is the posterior predictive mean. Black, solid circles denote unobserved data that fell within the 95\% point-wise posterior predictive intervals. Black `X's denote unobserved data that fell outside the 95\% point-wise posterior intervals. The empirical 95\%  coverage for all forecasts in 2015 was 95.2\%.}
\label{fig:fcst2015}
\end{figure}

Models 2015.3 through 2015.30, inclusively, were fit mimicking the sequential model fitting for an entire flu season. Weeks 3 and 30 roughly correspond to the forecasting window used in the CDC flu forecasting challenge. Figure \ref{fig:fcst2015} shows the posterior predictive mean and 95\% point-wise posterior predictive intervals for select model fits. Predictive uncertainty is largest when forecasts are made early in the flu season, but gradually diminishes throughout the flu season as more data are observed and incorporated into the model fitting. Forecasts early in the flu season reflect the bimodal nature of peak timing in the non-2015 flu seasons. That is, the forecasts for 2015 suggests there could either be an early peak to the flu season at week 13, or the peak could occur later around week 20. The average forecast exhibits a very sharp decline from week 13 to week 14, reflecting the decline in wILI from week 13 to week 14 in all non-2015 seasons (recall Figure \ref{fig:sirfit}). The reason the forecast exhibits bimodality is because of the hierarchical discrepancy model. Importantly, the empirical coverage for the 95\% nominal predictive intervals for all 2015.3 through 2015.30 model forecasts was 95.2\%. Empirical coverage will be discussed more in Section \ref{subsec:assessment}.

\begin{figure}[H]
\centering
\includegraphics[width=.9\textwidth]{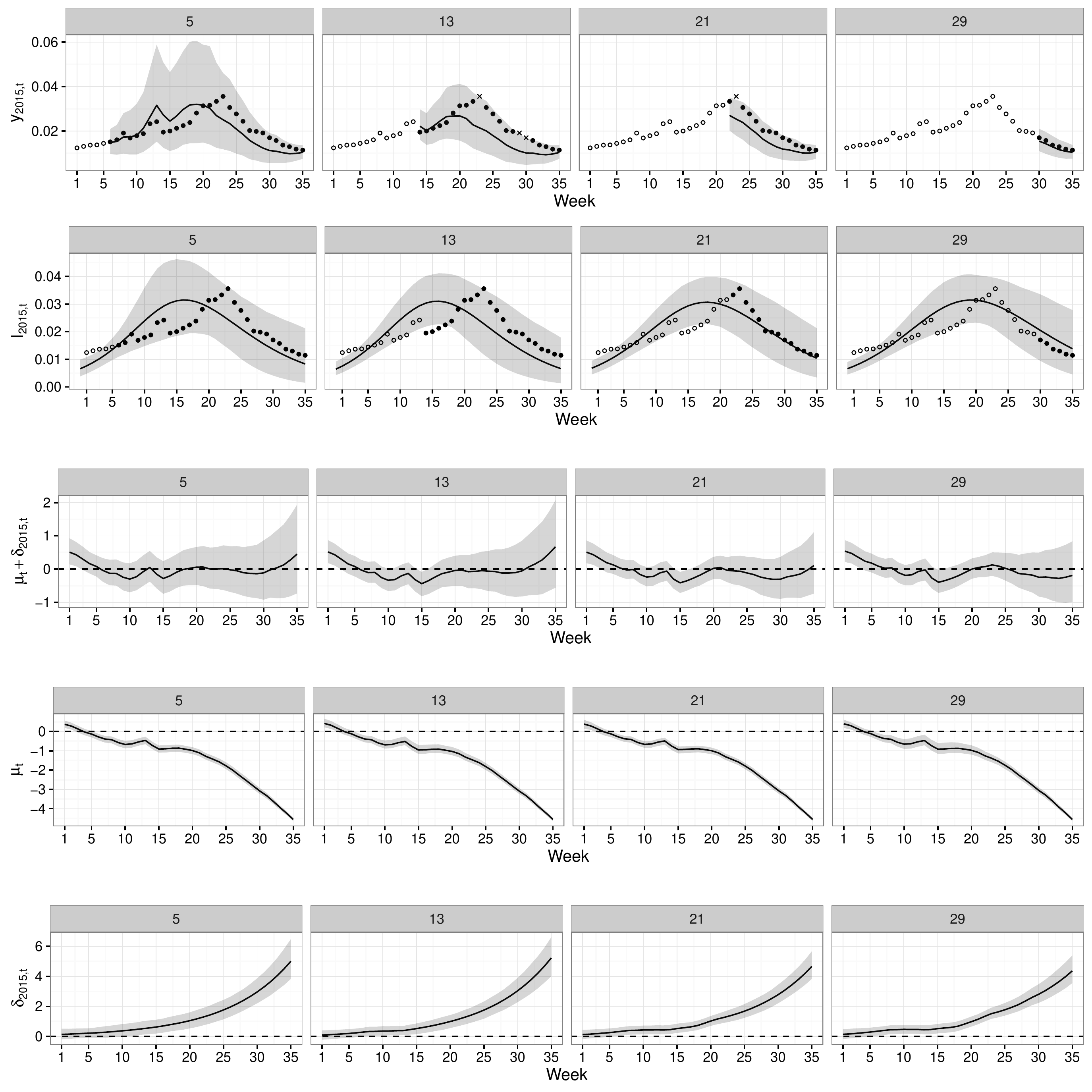}
\caption{Posterior predictive distribution for future wILI (top row) and posterior distributions for the various components of Equation \ref{eq:processmodel} (rows 2 through 5) for models 2015.5, 2015.10, 2015.15, 2015.20, 2015.25, and 2015.30 (columns from left to right). In the top row, hollow, black circles are observed wILI. Solid, black circles are unobserved wILI that fell within the 95\% predictive band. Black `X's are unobserved wILI that fell outside the 95\% predictive band. In the second row, hollow, black circles are observed wILI and solid, black circles are unobserved wILI. In all rows, the black line and grey bands represent the posterior mean and 95\% point-wise posterior interval for the row-specific quantity, respectively.}
\label{fig:modcomp2015}
\end{figure}

The 95\% point-wise posterior predictive intervals corresponding to the model components of Equation \ref{eq:processmodel} are presented in Figure \ref{fig:modcomp2015} for various weeks of the 2015 flu season. In the second row of Figure \ref{fig:modcomp2015}, we see the SIR component of the model approximates wILI. As more observations are incorporated into the analysis, the infectious trajectory better approximates the data. Though the infectious trajectory approximates the data, it is unable to match it exactly (as expected). Thus, there is a non-zero discrepancy, captured by $\mu_t + \delta_{2015,t}$ and plotted in the third row of Figure \ref{fig:modcomp2015}. When the average infectious trajectory underestimates the data, the discrepancy $\mu_t + \delta_{2015,t}$ is greater than zero and vice versa. The discrepancy, thus, compensates for the inadequacies of the infectious trajectory. The point-wise 95\% posterior interval for $\mu_t + \delta_{2015,t}$ is typically larger for weeks corresponding to unobserved data than weeks for observed data.

The discrepancy $\mu_t + \delta_{2015,t}$ is further decomposed into the common discrepancy $\mu_t$ and season-specific discrepancy $\delta_{2015,t}$ in the fourth and fifth rows of Figure \ref{fig:modcomp2015}, respectively. The posterior mean and 95\% point-wise posterior interval for $\mu_t$ is relatively unchanged for all model fits in Figure \ref{fig:modcomp2015}. This is because $\mu_t$ is common to all seasons, meaning the complete data from all 15 non-2015 flu seasons are informing the estimate of $\mu_t$. The incremental increase in data added to the analysis for season 2015 is a relatively small proportion of the total data informing $\mu_t$. We also see the estimate of $\mu_t$ captures the discrepancy bump between weeks 10 and 15. Finally, note that the posterior mean and 95\% posterior interval for $\mu_T$ is roughly -4.56 (-4.64, -4.47) for all models displayed in Figure \ref{fig:modcomp2015}. The mean of logit($y_{j,T}$) for all non-2015 seasons is -4.59. The identifying constraint of Equation \ref{eq:deltaprocessinit} effectively sets $\mu_T$ equal to the average of logit($y_{j,T}$), as supported by these results.

The season-specific discrepancy term, $\delta_{2015,t}$, does change throughout the season, as it is capturing the season-specific discrepancy unaccounted for by the common discrepancy and infectious trajectory. Also by the identifying constraint of Equation \ref{eq:deltaprocessinit}, $\delta_{2015,T}$ is set to -logit$(I_{2015,T})$. $\delta_{2015,t}$ gradually reverts to near zero from week $T$ to week 1, as was encouraged by Equations  \ref{eq:deltaprocess} and \ref{eq:alpha}.

\begin{figure}[!ht]
\centering
\includegraphics[width=.9\textwidth]{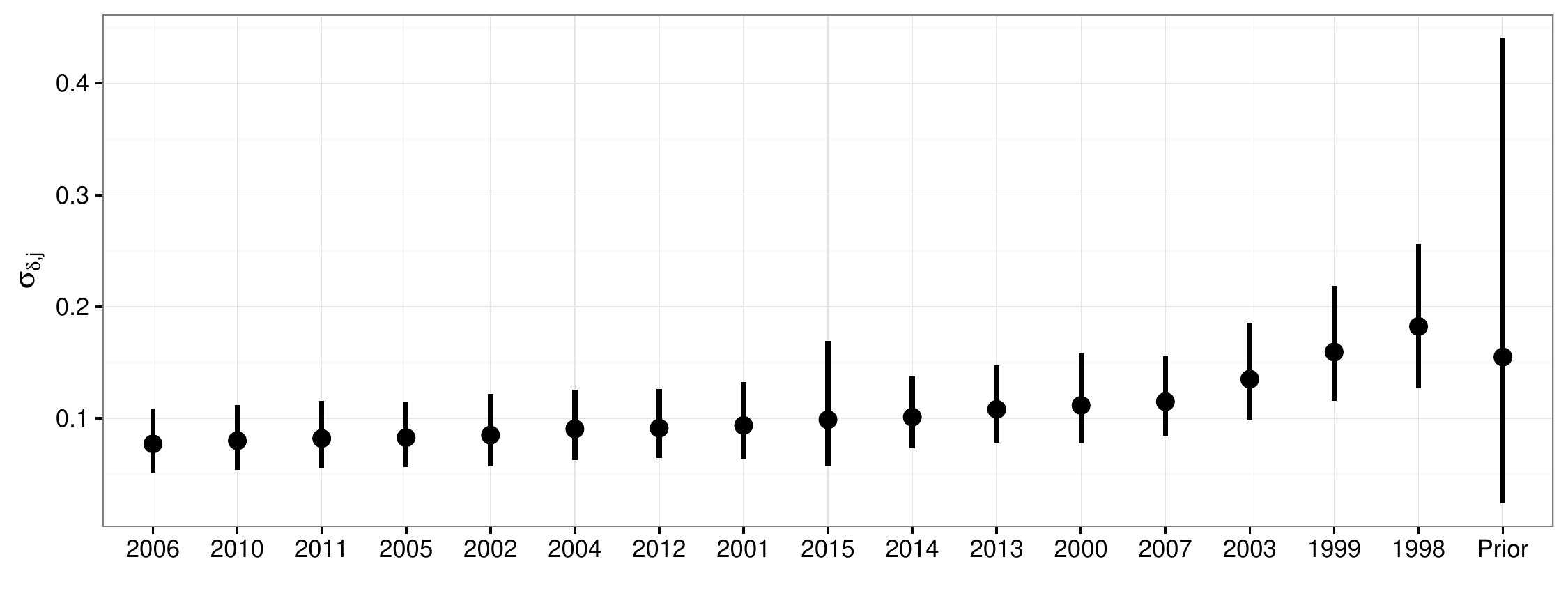}
\caption{95\% credible intervals (line segments) along with means (points) for the prior and posterior distributions for $\sigma_{\delta,j}$ for model 2015.3, with seasons ordered by descending posterior means.}
\label{fig:alphadelta}
\end{figure}

Figure \ref{fig:alphadelta} displays the posterior credible intervals for $\sigma_{\delta,j}$, the evolution standard deviation of the season-specific discrepancy trajectory. The larger the evolution standard deviation, the more flexible $\delta_{j,t}$ is. Greater flexibility for $\delta_{j,t}$ is needed for seasons whose wILI deviates more acutely from ``typical" wILI flu seasons. From Figure \ref{fig:alphadelta}, we see the three largest $\sigma_{\delta,j}$s as measured by posterior means correspond to flu seasons 1998, 1999, and 2003. These were also the three most ``atypical" flu seasons as measured by MSE in Figure \ref{fig:msewili}, supporting the interpretation that $\sigma_{\delta,j}$ captures season-specific discrepancy flexibility.

%
%
%
%

\subsection{Model Assessment}
\label{subsec:assessment}

For each of 16 flu seasons, we fit and forecast models Season.3 through Season.30. Recall that when we fit, for example, model 2015.3, we forecast weeks 4 through 35 (32 forecasts) of season 2015. When we fit model 2015.30, we forecast weeks 31 through 35 (5 forecasts) of season 2015. For each complete season, we make 518 forecasts. Weighted ILI for weeks 34 and 35 are unavailable for seasons 1998 through 2001, thus 462 forecasts were made for those seasons. The totality of all forecasts across all ``leave-one-season-out" model fits was 8,064 forecasts. Each forecast is a 95\% point-wise posterior predictive interval for a future observation of wILI. The overall empirical coverage was 89.4\%, suggesting the DB model forecasts, on balance, exhibit undercoverage.

\begin{figure}
    \centering
    \begin{subfigure}[b]{.4\textwidth}
        \includegraphics[width=1\textwidth]{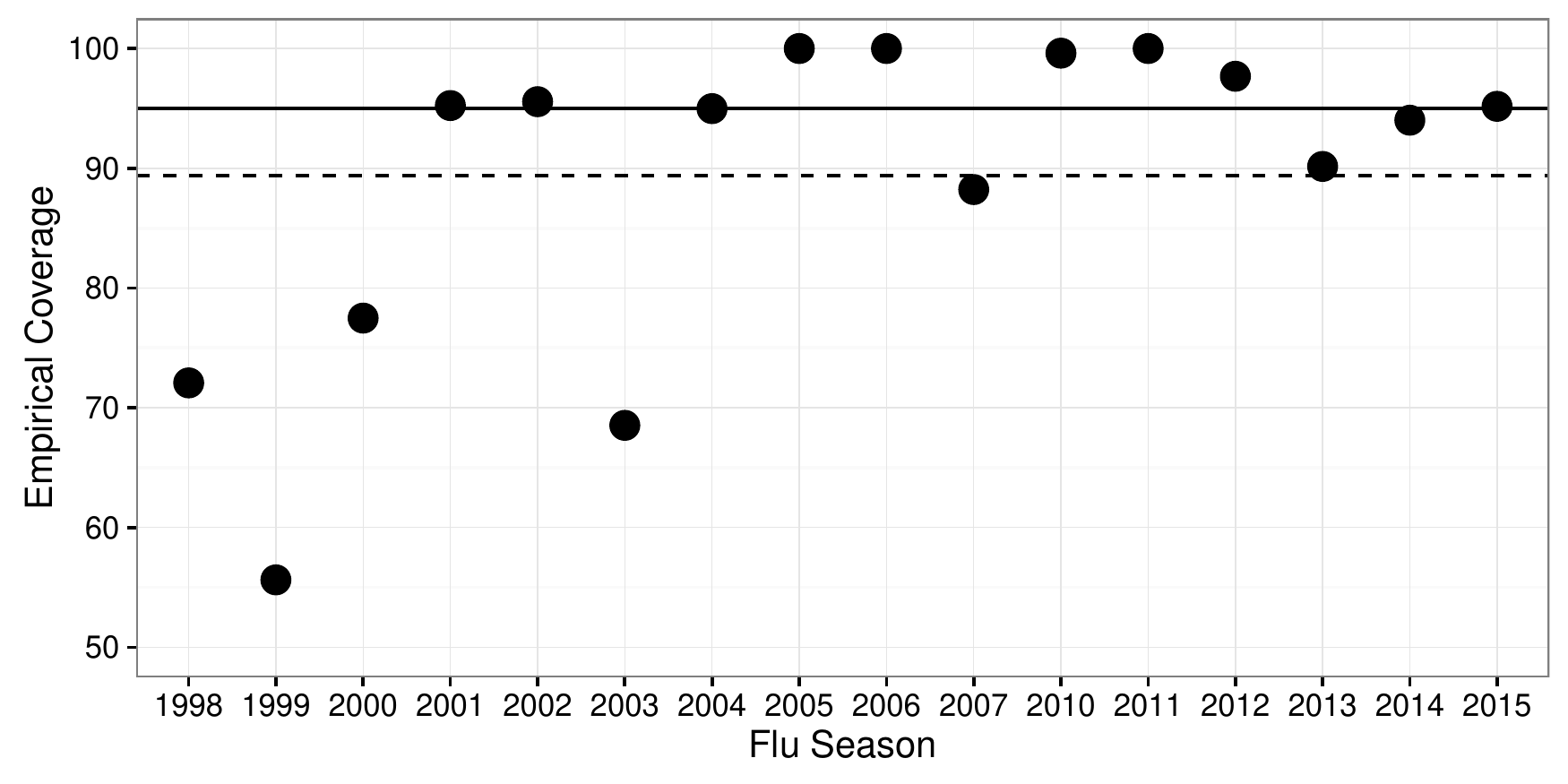}
        \caption{Flu season }
        \label{fig:coveragea}
    \end{subfigure}
    ~ 
    \begin{subfigure}[b]{.4\textwidth}
        \includegraphics[width=1\textwidth]{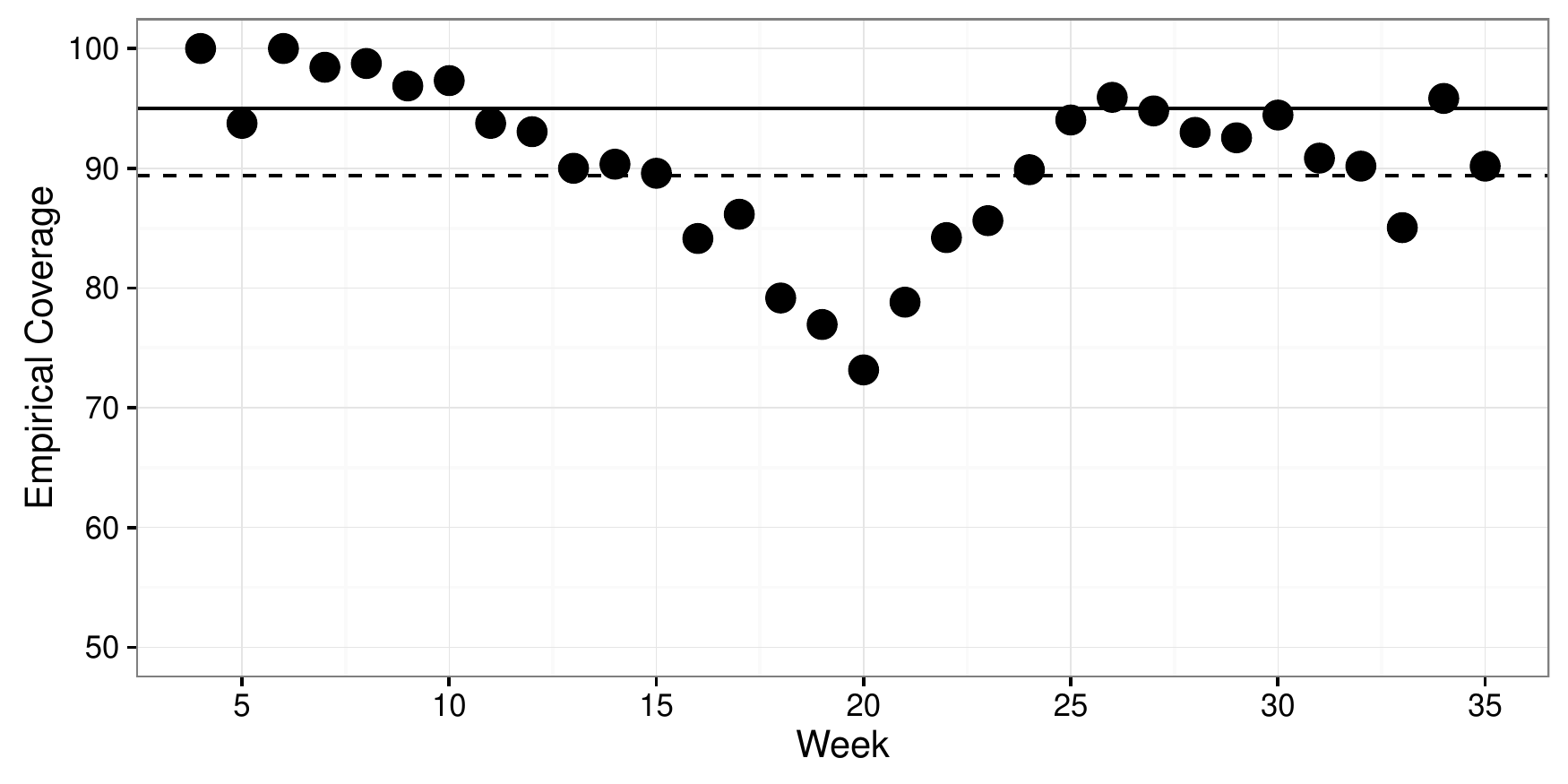}
        \caption{Week of flu season}
        \label{fig:coverageb}
    \end{subfigure}
    ~ 
        \begin{subfigure}[b]{.4\textwidth}
        \includegraphics[width=1\textwidth]{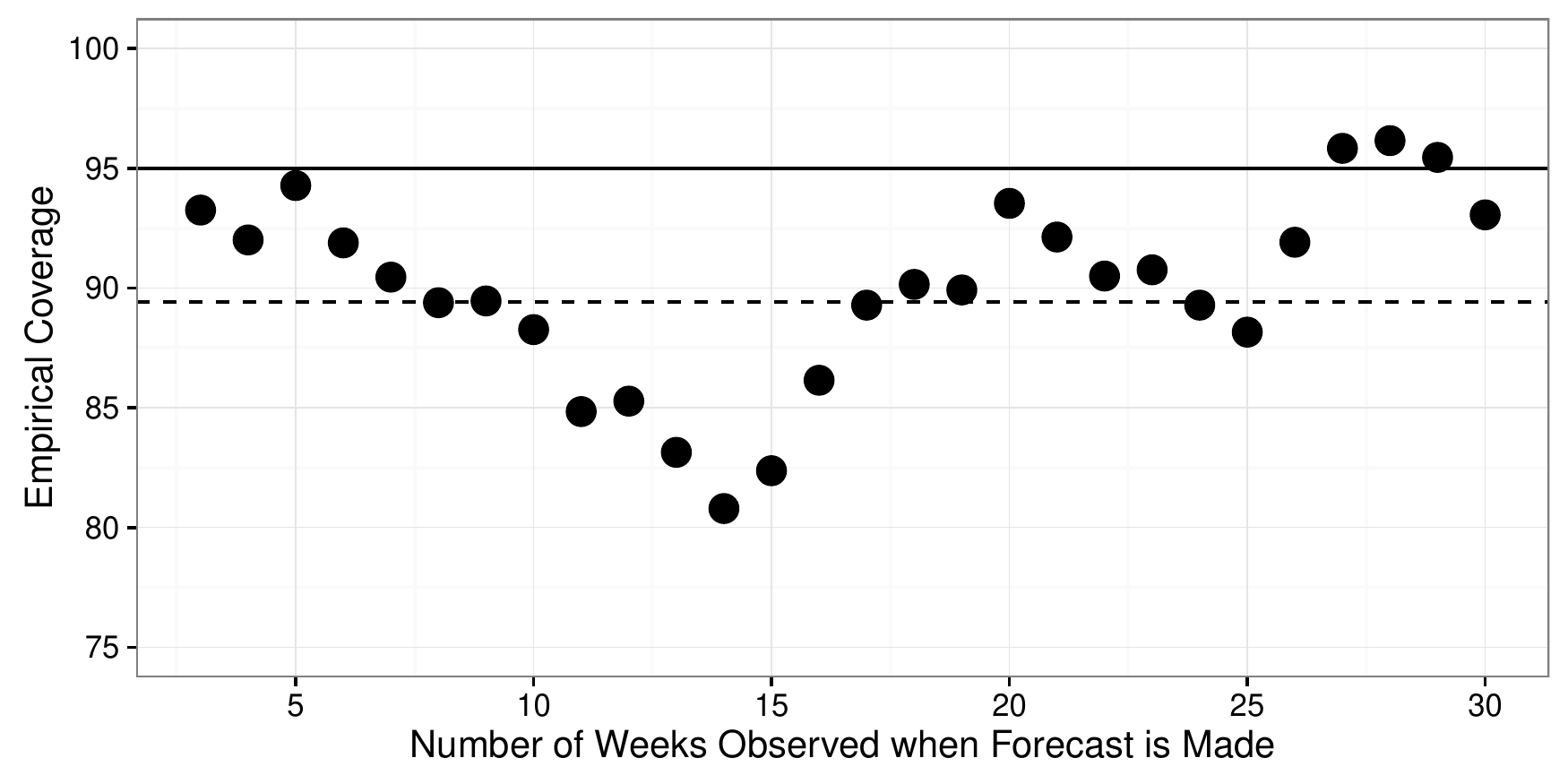}
        \caption{``Week" of ``Season.Week" model fit}
        \label{fig:coveragec}
    \end{subfigure}
    ~ 
    \begin{subfigure}[b]{.4\textwidth}
        \includegraphics[width=1\textwidth]{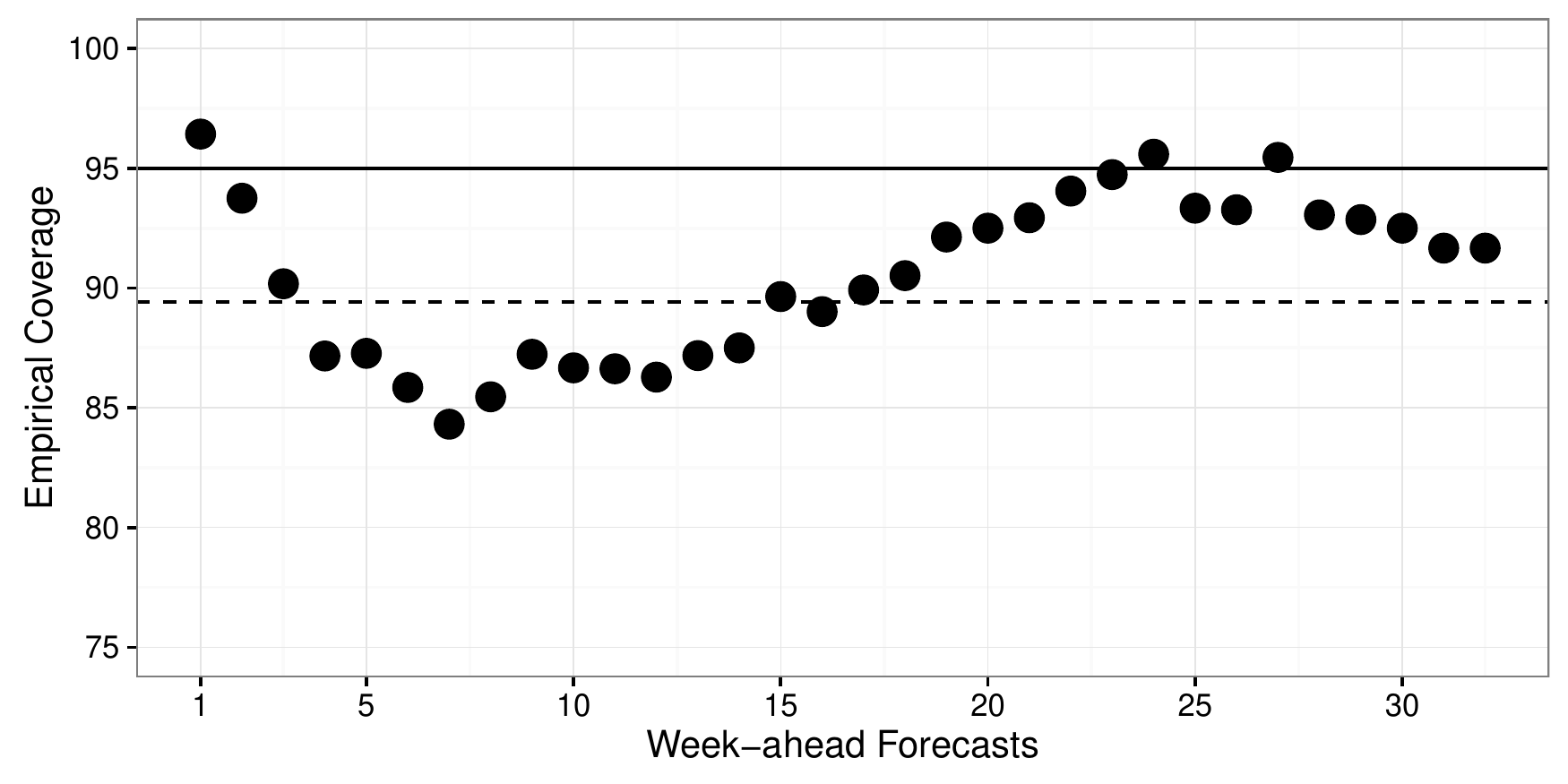}
        \caption{Week-ahead forecast.}
        \label{fig:coveraged}
    \end{subfigure}
    \caption{Empirical forecast coverage (points) by various partitionings of forecasts. The overall empirical coverage was 89.4\% (dashed line). The solid line represents the nominal 95\% coverage.}
    \label{fig:coverage}
\end{figure}

Figure \ref{fig:fcst2015} suggests that empirical coverage is correlated, with future wILI often either falling within or outside predictive intervals on consecutive weeks. We can interrogate the DB model's forecasting accuracy by subsetting the 8,064 forecasts. Figure \ref{fig:coverage} plots empirical coverage versus various partitionings of the forecasts, revealing both areas the DB model's forecasts perform well and need improvement as well as times in the forecasting process that are inherently more difficult to forecast than others.

Figure \ref{fig:coveragea} plots empirical coverage by flu season. We see empirical coverage ranges from 100\% during seasons 2005, 2006, and 2011 to 55.6\% in 1999. The flu seasons whose empirical coverage deviated most significantly below the nominal 95\% coverage were also the most ``atypical" flu seasons as determined by MSE in Figure \ref{fig:msewili}: 1998, 1999, and 2003. Hierarchical models are powerful models for borrowing strength across a collection of similar units (e.g., flu seasons). An underlying assumption of hierarchical models, however, is that a new flu season is ``similar" to the flu seasons that have already been seen (i.e., a new flu season comes from the same superpopulation of flu seasons the observed flu seasons came from). When a new flu season deviates from ``typical", forecasts struggle.   If seasons 1998, 1999, and 2003 were removed from the coverage assessment, the overall coverage would be 94.6\%. The forecast undercoverage appears to be driven by a few flu seasons, suggesting more variability should be accounted for in the model.

Figure \ref{fig:coverageb} plots empirical coverage by the week of the flu season. Empirical coverage ranged from 100\% on weeks 4 and 6 to 73.2\% on week 20. We see a general trend of declining empirical coverage from week 4 to week 20, followed by a recovery in coverage to week 35. Weeks near week 20 represent the most challenging period of the flu season to forecast, as wILI for some of the later peaking flu seasons are still ascending to their peak, while the early peaking flu seasons have been reverting to low levels for multiple weeks. Though the range in wILI at week 20 is the same as at week 13 (both 0.055, see Figure \ref{fig:wili} for reference), all wILI trajectories are ascending to week 13 while some are ascending and some are descending to week 20. This ascending/descending distinction is an added source of uncertainty for week 20 not shared with week 13, making it so challenging to forecast.

Figure \ref{fig:coveragec} plots the ``Week" of the Season.Week model fit. For example, the plotted empirical coverage for ``Week" 5 is the average empirical coverage over all forecasts made when only the first 5 weeks of the flu season have been observed. The empirical coverage ranges from 96.2\% for Season.28 models to 80.8\% for Season.14 models. There is a general decline in empirical coverage in Season.5 through Season.14, with an upturn in empirical coverage from Season.14 to Season.30. Forecasts for model Season.14 represent a fork in the forecasting process. Recall all flu seasons exhibit a downturn in wILI from week 13 to week 14. That downturn either signifies the worst of the flu season has occurred or a temporary decline in an otherwise still ascending flu season. Forecasts corresponding to model Season.14, thus, exhibit a lot of uncertainty. The information in wILI for weeks 15, 16, and 17 provide much information about whether the flu season will continue to ascend or descend.

Finally, Figure \ref{fig:coveraged} plots the empirical forecast versus the week-ahead-forecast. Empirical coverage ranges from 96.4\% for one-week-ahead forecasts to 84.3\% for seven-week-ahead forecasts. Empirical coverage generally declines from one to seven-week-ahead forecasts, and then increases with increasing week-ahead-forecasts. The average empirical coverage for all one to four-week ahead forecasts is 91.9\%, representing an improvement when compared to the overall empirical coverage of 89.4\%. Seasonal flu forecasting represents an example of forecasting where forecast accuracy does not decline with increasing week-ahead-forecasts. Figure \ref{fig:wili} illustrates why this is. Again, wILI is relatively well-behaved and predictable at the beginning and the end of the flu season. The bulk of the uncertainty occurs in the middle of the flu season. In October, it is easier to predict wILI in May (roughly 30 weeks into the future) than it is to predict wILI in December (roughly 10 weeks into the future).

\subsection{Model Comparison}
\label{subsec:comparison}

Forecasting challenges are effective ways to both identify and improve predictive capabilities in a myriad of fields, including influenza forecasting \citep{tetlock2017bringing}. The CDC has hosted an influenza forecasting challenge, open to the public, since the 2013\textendash 2014 flu season. In the inaugural 2013\textendash 2014 challenge, over a dozen forecasting models participated in the challenge \citep{biggerstaff2016results}. Since then, the challenge has grown, with 29 forecasting models participating in the 2016\textendash 2017 challenge \citep{cdcchallenge}. The CDC's flu forecasting challenge is an opportunity for the CDC to scope flu forecasting capabilities. It is also an opportunity for teams to compare their forecasting models against the leading forecasting models in the field. Competition drives innovation and incentives iterative improvement.

To see how the DB model compares to cutting-edge forecasting competition, we compare the DB model to the 14 models that participated in the 2015\textendash 2016 flu forecasting challenge. These models represent a diverse collection of mechanistic, machine learning, and statistical models making use of numerous data sources including Internet based sources such as Google, Wikipedia, and Twitter. The weekly submissions for all 14 models are publicly available \citep{pubsub}.

Model comparison follows the evaluation criteria of the 2015\textendash 2016 flu forecasting challenge \citep{contest2015evaluation}, comparing each model's ability to accurately predict seven targets throughout the flu season:
\begin{itemize}
\item \textbf{Peak intensity (PI)}: the maximum value of wILI for the flu season.
\item \textbf{Timing of peak intensity (PT)}: the week the PI occurs.
\item \textbf{Onset}: the start of the flu season, defined as the first of three consecutive weeks of wILI above the national baseline \citep{contest2015}. For the 2015\textendash 2016 flu season, the national baseline was defined as 0.021.
\item \textbf{One, two, three, and four week ahead forecasts}: short-term forecasts.
\end{itemize}

Each week of the flu season, a submission for all targets is made in the form of a probabilistic forecast. For each target and submission week, probabilities are assigned to mutually exclusive and exhaustive bins such that the probabilities sum to one. For PI and the short-term forecasts, bins range from 0 to 0.13, with bin widths of 0.005. For completion, there is a catch-all bin from 0.13 to 1. Thus, for PI and short-term forecasts, the probability space is partitioned into 27 bins. For PT and onset, each bin corresponds to a week. There is an additional bin of ``no onset" for onset, as there is no guarantee a flu season will have three consecutive weeks of wILI above baseline.

The evaluation of each target is done by computing a logarithmic score. Let
\begin{align}
\label{eq:pvec}
\boldsymbol{p}_{\text{mod,wk,tgt}} &= (p_{\text{mod,wk,tgt,1}},p_{\text{mod,wk,tgt,2}},\ldots,p_{\text{mod,wk,tgt},n_{\text{tgt}}})'
\end{align}
\noindent represent the vector of binned probabilities for model ``mod" submitted on week ``wk" corresponding to target ``tgt", where $p_{\text{mod,wk,tgt},i}$ corresponds to the probability assigned to bin $i = 1,2,\ldots,n_{\text{tgt}}$.

Assume the target falls in bin $i^*$. Then, the log score corresponding to $\boldsymbol{p}_{\text{mod,wk,tgt}}$ is defined as,
\begin{align}
\label{eq:logscore}
S(\boldsymbol{p}_{\text{mod,wk,tgt}},i^*) &= ln(p_{\text{mod,wk,tgt},i^*-1} + p_{\text{mod,wk,tgt},i^*} + p_{\text{mod,wk,tgt},i^*+1}).
\end{align}
\noindent That is, the log score is the natural log of the sum of the probabilities assigned to the correct bin and the immediately preceding and proceeding bins. For example, if the forecasted probability of the PT occurring on weeks 19, 20, and 21 are 0.1, 0.3, and 0.2, respectively, and the true PT is week 20, then the log score is $ln(0.1 + 0.3 + 0.2) = -0.51$. A perfect log score is 0 and is achieved if all the probability is assigned to the correct or immediately adjacent bins. A score of -10 is assigned to all undefined natural log scores (e.g., $ln(0)$ is undefined), all late submissions, and all submissions where the sum of $\boldsymbol{p}_{\text{mod,wk,tgt}} > 1.1$. Good forecasts, as determined by Equation \ref{eq:logscore}, are forecasts that concentrate appreciable probability tightly around the bin of the true target value.

For the 14 participating models, the binned probabilities are publicly available. For the DB model, bins were populated by simulating from the posterior predictive distribution for each target and submission week.

Weighted ILI estimates are revised every week. For example, when wILI on week 3 of the 2015\textendash 2016 flu season was first publicly released, it was 0.0135. The next week, when week 4 was first publicly released, the estimate for wILI on week 3 was revised from 0.0135 to 0.0141. This process of revision can occur every week. These weekly revisions can cause wrinkles when retrospectively comparing a new model (e.g., the DB model) to models that participated in past forecasting challenges, as the retrospective model fitting is often based on wILI estimates that were unavailable on the date real-time submissions were made. For the comparison of the DB model to the other models, we used the wILI estimates that were available for each submission week of the 2015\textendash 2016 forecasting challenge allowing us to faithfully replicate the forecasting conditions. In general, though, the weekly wILI revisions can cause problems with retrospective model comparisons.

\begin{figure}[!ht]
\centering
\includegraphics[width=.5\textheight]{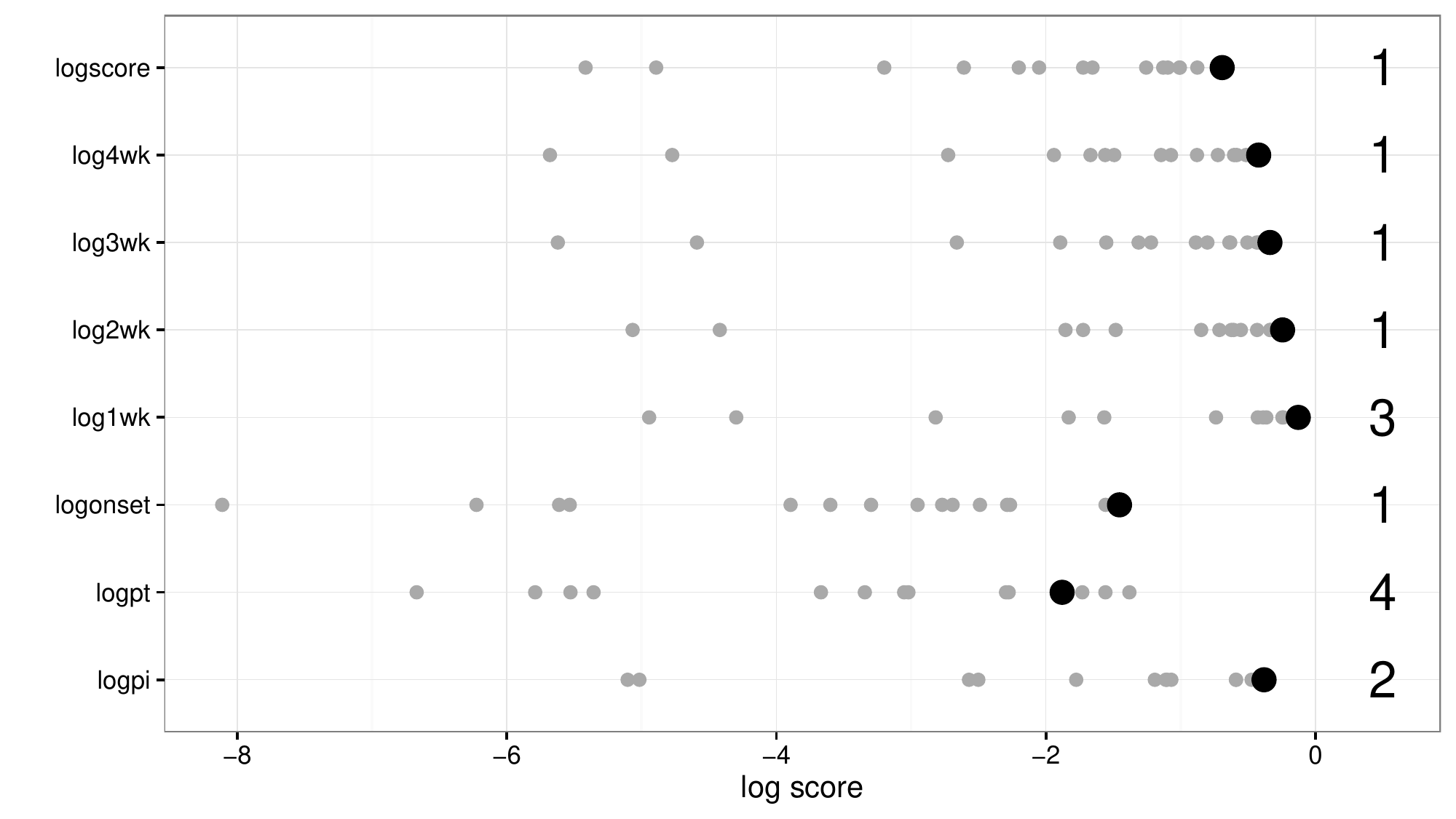}
\caption{Average log score for every target and model in the CDC's 2015\textendash 2016 flu forecasting competition (grey) and the DB model (black). The overall average log score is shown as `logscore'. The numbers on the right represent the rank of the DB model relative to all other models where 1 is the best log score. The DB model had the best overall average log score.}
\label{fig:comparison}
\end{figure}

For every target and model, we computed the average log score over all submission weeks. The results are plotted in Figure \ref{fig:comparison}. The DB model compared favorably to all other models with respect to all targets. The DB model beat all other models with respect to onset and two through four week ahead forecasts and ranked no worse than fourth for all targets. Averaging over the log scores for all targets provides an estimate of a model's forecasting ability. The DB model beat all other models with respect to overall average log score.

As shown in Table \ref{tab:results}, the DB model was the only model to rank first for more than one target, indicating that it was the best forecasting model with respect to multiple targets. The DB model was also the only model to rank no worse than fourth with respect to all forecasting targets, suggesting the DB model was not deficient at forecasting any flu season target, in a relative sense. Five comparison models did not rank first for any targets but ranked no worse than fourth for at least one target, while six models ranked worse than fourth for all targets.

\begin{table}[ht]
\centering
\caption{The number of top 1 through top 4 rankings by each model. The DB model was the only model to have more than one top 1 ranking and the only model to rank no worse than fourth for all seven targets. Model 1 through Model 14 represent the anonymized comparison models.}
\label{tab:results}
\begin{tabular}{ccccc}
  \hline
\textbf{Model} & \textbf{Top 1} & \textbf{Top 2} & \textbf{Top 3} & \textbf{Top 4} \\
  \hline
DB &   4 &   5 &   6 &   7 \\
Model 1 &   1 &   4 &   4 &   5 \\
Model 2 &   1 &   1 &   1 &   2 \\
Model 3 &   1 &   1 &   1 &   1 \\ \hline
Model 4 &   0 &   2 &   2 &   2 \\
Model 5 &   0 &   1 &   3 &   6 \\
Model 6 &   0 &   0 &   2 &   3 \\
Model 7 &   0 &   0 &   1 &   1 \\
Model 8 &   0 &   0 &   1 &   1 \\ \hline
Models 9 \textendash  14 &   0 &   0 &   0 &   0 \\
   \hline
\end{tabular}
\end{table}

\begin{figure}[!ht]
\centering
\includegraphics[width=.7\textwidth]{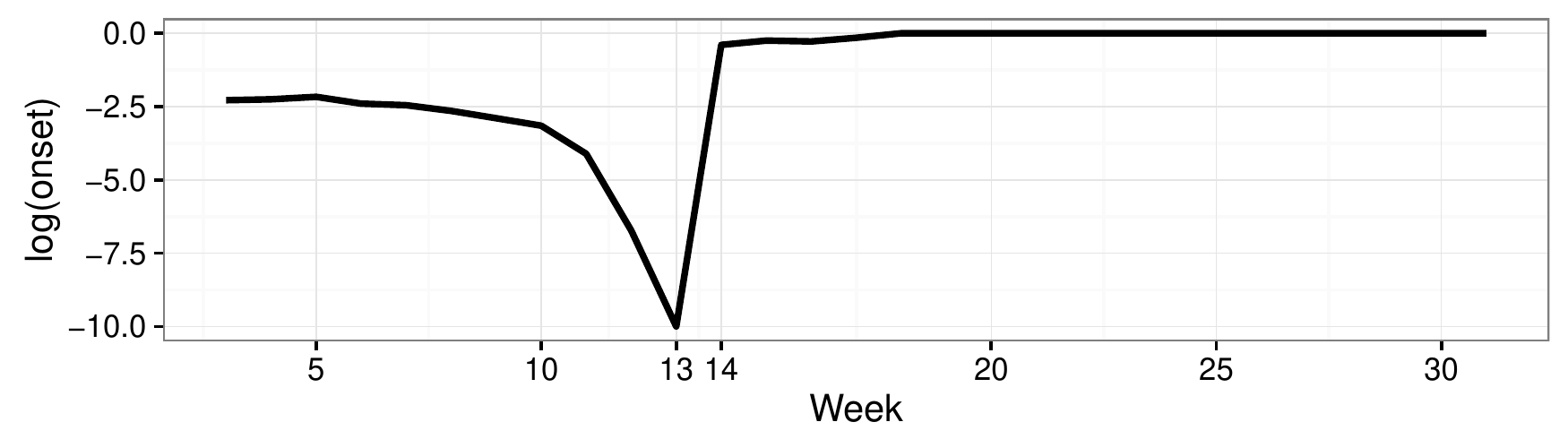}
\includegraphics[width=.7\textwidth]{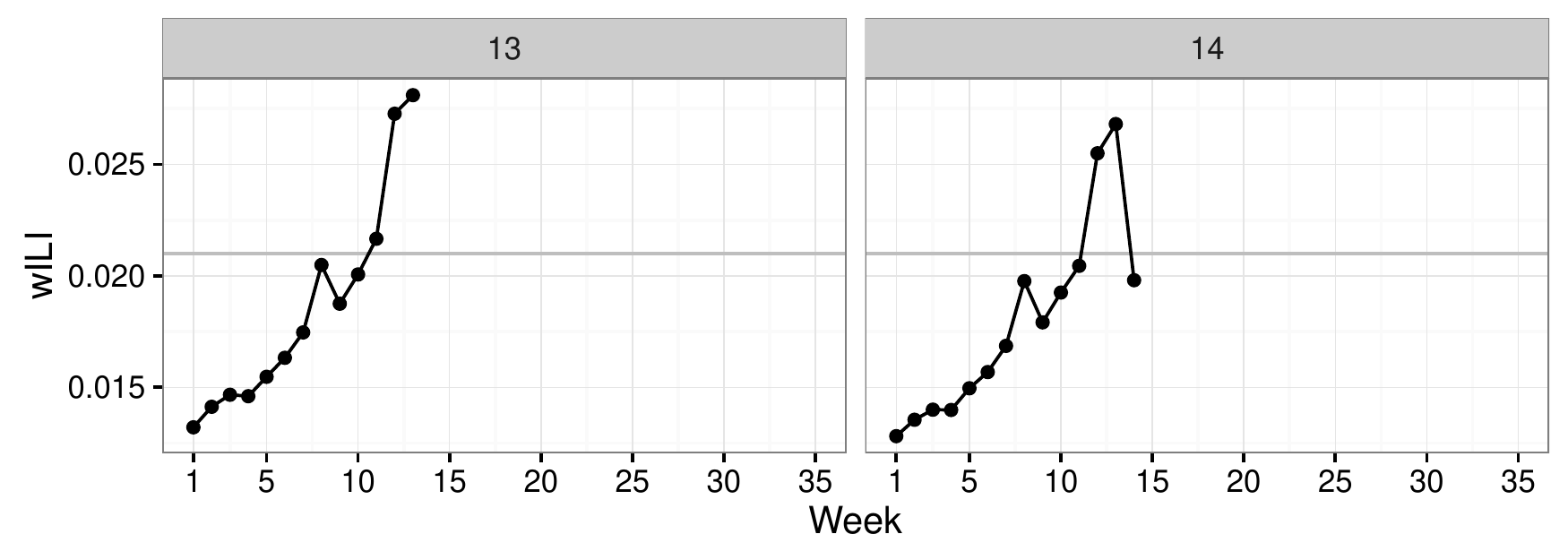}
\caption{(Top) The weekly onset log score for the DB model. (Bottom) Weighted ILI estimates available through the first 13 and 14 weeks of the 2015\textendash 2016 flu season, respectively. The grey, horizontal line is the national baseline of 0.021.}
\label{fig:onsettarget}
\end{figure}

Though the DB model had the best overall log score and appears to be one of the best models for all considered targets, there is still room for improvement. Figure \ref{fig:onsettarget} illustrates one such area and articulates why accounting for the weekly wILI revisions is important for faithful retrospective comparisons between new models and flu challenge participating models. The top of Figure \ref{fig:onsettarget} shows that on week 13, the DB's log score for onset was -10, indicating zero probability was assigned to the correct or neighboring bins of the true onset (week 16). The bottom of Figure \ref{fig:onsettarget} reveals why this occurred. Based on the data available through the first 13 weeks of the 2015\textendash 2016 flu season, week 11 was, by definition, the onset (i.e., week 11 was the first of three consecutive weeks above baseline). Thus, on week 13, the DB model forecasted week 11 to be the onset with probability one. Declaring week 11 the onset, however, ignores the weekly wILI revisions. The very next week when 14 weeks of wILI estimates were available, week 11's wILI estimate was revised and fell below the national baseline, indicating it was not the onset. The scenario displayed in Figure \ref{fig:onsettarget} articulates that the forecasts of the DB model are missing a source of uncertainty caused by wILI revisions. It also articulates that using revised versions of wILI estimates to retrospectively compare a model to forecasts based on currently available wILI estimates gives an unfair advantage to the new model. The log onset score of -10 would not have occurred for the DB model were forecasts based on wILI estimates available at the end of the flu season, biasing the log scores up.

Even without accounting for the uncertainty caused by weekly wILI revisions in the forecasts, the DB model outperformed all models it was compared against, suggesting that it is one of the leading flu forecasting models with room for improvement.

\section{Discussion}
\label{sec:discussion}

In this paper, we introduced a novel dynamic Bayesian influenza forecasting model that exploits discrepancy structure. The basic insight and motivation leading to the development of the DB model is that the disease transmission model (e.g., the SIR model) and the data-generating model are not equivalent; disease transmission is a component of but not equal to the data-generating process. The data-generating model is non-exhaustively comprised of a disease transmission process, a healthcare provider visitation process, an influenza-like illness determination process, and a reporting process. Thus, even if a disease transmission model more sophisticated than the SIR model were used, of which there are numerous (e.g., the SIRS model, the SEIR model), there might still be a disagreement between the best version of the disease transmission model and the data. Rather than attempt to model each component of the data-generating model, we acknowledge there will likely be a systematic disagreement between the best version of the disease transmission model and the data. We then model the commonalities of the discrepancy across flu seasons with a flexible, hierarchical model. The hierarchical discrepancy model allows us to leverage patterns in the data the disease transmission model is incapable of capturing and not simply model the discrepancy as white noise.

The DB model assumes future flu seasons will exhibit similar trajectories to past flu seasons. We showed that the more dissimilar a flu season was as compared to the other considered flu seasons, the worse forecasts were. Because of this underlying assumption, the DB model would be inappropriate for forecasting pandemic influenza. A companion model to the DB model that tracks how ``dissimilar" a new flu season is compared to the collection of observed seasonal flu seasons could be useful to gauge when the DB model should be trusted and when it should not.

When compared to the forecasting models participating in the CDC's 2015\textendash 2016 flu forecasting competition, the DB model had the best overall score, was the only model to rank first with respect to multiple targets, and never ranked worse than fourth with respect to all forecasting targets. Comparisons were facilitated by the CDC coordinating the flu forecasting challenge and making the submissions publicly available. These submissions provide an excellent test case for future models to be compared against.

The work of \cite{ginsberg2009detecting} demonstrated the potential value of monitoring flu outbreaks with Google search queries. The basic idea being, when individuals experience symptoms of the flu, they may go to their web browser to search for more information. Thus, an increase in searches for flu related terms may indicate an increase in flu incidence in the population. The work of \cite{ginsberg2009detecting} sparked a large research effort to investigate other digital surveillance sources and their possible connection to disease surveillance \citep[e.g.,][]{generous2014global,wilson2009early,polgreen2008using}. Many forecasting models have augmented wILI with digital surveillance data \citep[e.g.,][]{hickmann2015forecasting,roni2015,shaman2013}, including those that participated in the 2013\textendash 2014 CDC flu forecasting competition \citep{biggerstaff2016results}. Recently, the value of digital surveillance data with respect to flu forecasting has been curbed \citep[e.g.,][]{lazer2014parable,priedhorsky2017measuring}. In fact, as \cite{biggerstaff2016results} conclude from the 2013\textendash 2014 flu forecasting competition, ``not all digital data are equally accurate, and the algorithms and methodologies underpinning these data require constant upkeep to maintain their accuracy. \ldots Influenza forecasting models informed by digital data are subject to the biases and errors of their underlying source data." The DB model does not use digital surveillance data. Incorporating digital surveillance data may or may not improve forecasts; investigation into this might serve as a next iteration of the DB model. It is worth noting, however, that even without digital surveillance data, the DB model compared favorably to all comparison models, some of which did make use of digital surveillance data. The results presented in Section \ref{subsec:comparison} suggest influenza forecasting can be improved without augmenting wILI with digital surveillance data but rather focusing on statistical model development.

\begin{appendices}
\section{Fourth order Runge-Kutta approximation method}
\label{appendix:rk4}

The SIR model does not have an explicit, closed form solution. We approximate the solution to the SIR model with the fourth order Runge-Kutta (RK4) approximation to Equation \ref{eq:sir}. We recursively define $S_{j,t}$, $I_{j,t}$, and $R_{j,t}$ as described in Equation \ref{eq:rk4}:
\begin{align}
\label{eq:rk4}
\begin{pmatrix}
S_{j,t}\\
I_{j,t}\\
R_{j,t}\\
\end{pmatrix}&=
\begin{pmatrix}
S_{j,t-1} + \frac{1}{6}[k^{S1}_{j,t-1} + 2k^{S2}_{j,t-1} + 2k^{S3}_{j,t-1} + k^{S4}_{j,t-1}] \\
I_{j,t-1} + \frac{1}{6}[k^{I1}_{j,t-1} + 2k^{I2}_{j,t-1} + 2k^{I3}_{j,t-1} + k^{I4}_{j,t-1}]\\
R_{j,t-1} + \frac{1}{6}[k^{R1}_{j,t-1} + 2k^{R2}_{j,t-1} + 2k^{R3}_{j,t-1} + k^{R4}_{j,t-1}]
\end{pmatrix},
\end{align}

\noindent where
\begin{subequations}
\label{eq:k}
\begin{align}
\begin{split}
\label{eq:ks}
k^{S1}_{j,t-1} &= -\beta_j   S_{j,t-1} I_{j,t-1},\\
k^{S2}_{j,t-1} &= -\beta_j  [S_{j,t-1}+.5k^{S1}_{j,t-1}][I_{j,t-1}+.5k^{I1}_{j,t-1}], \\
k^{S3}_{j,t-1} &= -\beta_j  [S_{j,t-1}+.5k^{S2}_{j,t-1}][I_{j,t-1}+.5k^{I2}_{j,t-1}], \\
k^{S4}_{j,t-1} &= -\beta_j  [S_{j,t-1}+ k^{S3}_{j,t-1}][I_{j,t-1}+k^{I3}_{j,t-1}],
\end{split}\\ \nonumber \\
\begin{split}
\label{eq:ki}
k^{I1}_{j,t-1} &= \beta_j   S_{j,t-1}I_{j,t-1} - \gamma_j I_{j,t-1},\\
k^{I2}_{j,t-1} &= \beta_j  [S_{j,t-1}+.5k^{S1}_{j,t-1}][I_{j,t-1}+.5k^{I1}_{j,t-1}] - \gamma_j [I_{j,t-1}+.5k^{I1}_{j,t-1}],\\
k^{I3}_{j,t-1} &= \beta_j  [S_{j,t-1}+.5k^{S2}_{j,t-1}][I_{j,t-1}+.5k^{I2}_{j,t-1}] - \gamma_j [I_{j,t-1}+.5k^{I2}_{j,t-1}],\\
k^{I4}_{j,t-1} &= \beta_j  [S_{j,t-1}+k^{S3}_{j,t-1}][I_{j,t-1}+k^{I3}_{j,t-1}] - \gamma_j [I_{j,t-1}+k^{I3}_{j,t-1}],
\end{split}\\ \nonumber \\
\begin{split}
\label{eq:kr}
k^{R1}_{j,t-1} &= \gamma_j  I_{j,t-1}, \\
k^{R2}_{j,t-1} &= \gamma_j [I_{j,t-1}+.5k^{I1}_{j,t-1}],\\
k^{R3}_{j,t-1} &= \gamma_j [I_{j,t-1}+.5k^{I2}_{j,t-1}], \\
k^{R4}_{j,t-1} &= \gamma_j [I_{j,t-1}+k^{I3}_{j,t-1}].
\end{split}\\ \nonumber
\end{align}
\end{subequations}

\end{appendices}

\section*{Acknowledgements}
This work is supported in part by NIH/NIGMS/MIDAS under grant U01-GM097658-01 and by the U.S. Department of Energy, Office of Science, Advanced Scientific Computing Research SciDAC program. LANL is operated by Los Alamos National Security, LLC for the Department of Energy under contract DE-AC52-06NA25396. The funders had no role in study design, data collection and analysis, decision to publish, or preparation of the manuscript. Approved for public release: LA-UR-17-22749

\bibliography{dynamicbayes.bib}

\begin{thebibliography}{49}
\providecommand{\natexlab}[1]{#1}
\providecommand{\url}[1]{\texttt{#1}}
\expandafter\ifx\csname urlstyle\endcsname\relax
  \providecommand{\doi}[1]{doi: #1}\else
  \providecommand{\doi}{doi: \begingroup \urlstyle{rm}\Url}\fi

\bibitem[Bardak and Tan(2015)]{bardak2015prediction}
Batuhan Bardak and Mehmet Tan.
\newblock Prediction of influenza outbreaks by integrating {W}ikipedia article
  access logs and {G}oogle flu trend data.
\newblock In \emph{Bioinformatics and Bioengineering (BIBE), 2015 IEEE 15th
  International Conference on}, pages 1--6. IEEE, 2015.

\bibitem[Bayarri et~al.(2007)Bayarri, Berger, Paulo, Sacks, Cafeo, Cavendish,
  Lin, and Tu]{bayarri2007}
Maria~J. Bayarri, James~O. Berger, Rui Paulo, Jerry Sacks, John~A. Cafeo, James
  Cavendish, Chin-Hsu Lin, and Jian Tu.
\newblock A framework for validation of computer models.
\newblock \emph{Technometrics}, 49\penalty0 (2):\penalty0 138--154, 2007.

\bibitem[Biggerstaff et~al.(2016)Biggerstaff, Alper, Dredze, Fox, Fung,
  Hickmann, Lewis, Rosenfeld, Shaman, Tsou, Velardi, Vespignani, Finelli, and
  {the Influenza Forecasting Contest Working Group}]{biggerstaff2016results}
Matthew Biggerstaff, David Alper, Mark Dredze, Spencer Fox, Isaac Chun-Hai
  Fung, Kyle~S. Hickmann, Bryan Lewis, Roni Rosenfeld, Jeffrey Shaman,
  Ming-Hsiang Tsou, Paola Velardi, Alessandro Vespignani, Lyn Finelli, and {the
  Influenza Forecasting Contest Working Group}.
\newblock Results from the centers for disease control and prevention’s
  predict the 2013--2014 {I}nfluenza {S}eason {C}hallenge.
\newblock \emph{BMC Infectious Diseases}, 16\penalty0 (1):\penalty0 357, 2016.

\bibitem[Brooks et~al.(2015)Brooks, Farrow, Hyun, Tibshirani, and
  Rosenfeld]{roni2015}
Logan~C. Brooks, David~C. Farrow, Sangwon Hyun, Ryan~J. Tibshirani, and Roni
  Rosenfeld.
\newblock Flexible modeling of epidemics with an empirical {B}ayes framework.
\newblock \emph{PLoS Comput Biol}, 11\penalty0 (8):\penalty0 e1004382, 2015.

\bibitem[Brynjarsd{\'o}ttir and {O'Hagan}(2014)]{brynjarsdottir2014learning}
Jenn{\`y} Brynjarsd{\'o}ttir and Anthony {O'Hagan}.
\newblock Learning about physical parameters: the importance of model
  discrepancy.
\newblock \emph{Inverse Problems}, 30\penalty0 (11):\penalty0 114007, 2014.

\bibitem[Capaldi et~al.(2012)Capaldi, Behrend, Berman, Smith, Wright, and
  Lloyd]{capaldi2012}
Alex Capaldi, Samuel Behrend, Benjamin Berman, Jason Smith, Justin Wright, and
  Alun~L. Lloyd.
\newblock Parameter estimation and uncertainty quantification for an epidemic
  model.
\newblock \emph{Mathematical Biosciences and Engineering}, page 553, 2012.

\bibitem[{Centers for Disease Control and
  Prevention}(2016{\natexlab{a}})]{cdcmmwr}
{Centers for Disease Control and Prevention}.
\newblock {MMWR} weeks, 2016{\natexlab{a}}.
\newblock URL \url{https://wwwn.cdc.gov/nndss/document/MMWR_week_overview.pdf}.
\newblock Accessed: 03-18-2017.

\bibitem[{Centers for Disease Control and
  Prevention}(2016{\natexlab{b}})]{cdcpandemic}
{Centers for Disease Control and Prevention}.
\newblock Past pandemics, 2016{\natexlab{b}}.
\newblock URL
  \url{https://www.cdc.gov/flu/pandemic-resources/basics/past-pandemics.html}.
\newblock Accessed: 03-18-2017.

\bibitem[{Centers for Disease Control and
  Prevention}(2016{\natexlab{c}})]{cdcsurveillance}
{Centers for Disease Control and Prevention}.
\newblock Overview of influenza surveillance in the {U}nited {S}tates,
  2016{\natexlab{c}}.
\newblock URL \url{https://www.cdc.gov/flu/weekly/overview.htm}.
\newblock Accessed: 03-18-2017.

\bibitem[Chretien et~al.(2014)Chretien, George, Shaman, Chitale, and
  McKenzie]{chretien2014influenza}
Jean-Paul Chretien, Dylan George, Jeffrey Shaman, Rohit~A. Chitale, and
  F.~Ellis McKenzie.
\newblock Influenza forecasting in human populations: a scoping review.
\newblock \emph{PloS one}, 9\penalty0 (4):\penalty0 e94130, 2014.

\bibitem[Coelho et~al.(2011)Coelho, Code{\c{c}}o, and
  Gomes]{coelho2011bayesian}
Flavio~Code{\c{c}}o Coelho, Claudia~Torres Code{\c{c}}o, and M.~Gabriela~M.
  Gomes.
\newblock A {B}ayesian framework for parameter estimation in dynamical models.
\newblock \emph{PloS one}, 6\penalty0 (5):\penalty0 e19616, 2011.

\bibitem[Dukic et~al.(2012)Dukic, Lopes, and Polson]{dukic2012tracking}
Vanja Dukic, Hedibert~F. Lopes, and Nicholas~G. Polson.
\newblock Tracking epidemics with {G}oogle flu trends data and a state-space
  {SEIR} model.
\newblock \emph{Journal of the American Statistical Association}, 107\penalty0
  (500):\penalty0 1410--1426, 2012.

\bibitem[{Epidemic Prediction Initiative}(2015{\natexlab{a}})]{contest2015}
{Epidemic Prediction Initiative}.
\newblock Forecast targets, 2015{\natexlab{a}}.
\newblock URL \url{https://predict.phiresearchlab.org/legacy/flu/targets.html}.
\newblock Accessed: 03-21-2017.

\bibitem[{Epidemic Prediction
  Initiative}(2015{\natexlab{b}})]{contest2015evaluation}
{Epidemic Prediction Initiative}.
\newblock Forecast evaluation, 2015{\natexlab{b}}.
\newblock URL
  \url{https://predict.phiresearchlab.org/legacy/flu/evaluation.html}.
\newblock Accessed: 03-21-2017.

\bibitem[{Epidemic Prediction Initiative}(2016)]{cdcchallenge}
{Epidemic Prediction Initiative}.
\newblock Flusight 2016-17: Seasonal influenza forecasting, 2016.
\newblock URL
  \url{https://predict.phiresearchlab.org/post/57f3f440123b0f563ece2576}.
\newblock Accessed: 03-18-2017.

\bibitem[Ewing et~al.(2016)Ewing, Lee, Viboud, and Bansal]{ewing2016contact}
Anne Ewing, Elizabeth~C. Lee, C{\'e}cile Viboud, and Shweta Bansal.
\newblock Contact, travel, and transmission: the impact of winter holidays on
  influenza dynamics in the {U}nited {S}tates.
\newblock \emph{Journal of Infectious Diseases}, page jiw642, 2016.

\bibitem[{FluSight Influenza Forecasting Challenge}(2016)]{pubsub}
{FluSight Influenza Forecasting Challenge}.
\newblock Ensemble forecasts for the {CDC}'s 2015\textendash 2016 {F}lu
  {F}orecasting {C}hallenge, 2016.
\newblock URL
  \url{https://github.com/cdcepi/FluSight_ensemble/tree/master/Data/2015-2016}.
\newblock Accessed: 04-18-2017.

\bibitem[Garza et~al.(2013)Garza, Basurto-D{\'a}vila, Ortega-Sanchez,
  Oreste~Carlino, Meltzer, Albalak, Balbuena, Orellano, Widdowson, and
  Averhoff]{garza2013effect}
Roberto~C. Garza, Ricardo Basurto-D{\'a}vila, Ismael~R. Ortega-Sanchez, Luis
  Oreste~Carlino, Martin~I. Meltzer, Rachel Albalak, Karina Balbuena, Pablo
  Orellano, Marc-Alain Widdowson, and Francisco Averhoff.
\newblock Effect of winter school breaks on influenza-like illness,
  {A}rgentina, 2005--2008.
\newblock \emph{Emerging Infectious Disease}, 19\penalty0 (6), 2013.

\bibitem[Gelman and Rubin(1992)]{gelman1992}
Andrew Gelman and Donald~B. Rubin.
\newblock Inference from iterative simulation using multiple sequences.
\newblock \emph{Statistical Science}, pages 457--472, 1992.

\bibitem[Generous et~al.(2014)Generous, Fairchild, Deshpande, Del~Valle, and
  Priedhorsky]{generous2014global}
Nicholas Generous, Geoffrey Fairchild, Alina Deshpande, Sara~Y. Del~Valle, and
  Reid Priedhorsky.
\newblock Global disease monitoring and forecasting with {W}ikipedia.
\newblock \emph{PLoS Computational Biology}, 10\penalty0 (11):\penalty0
  e1003892, 2014.

\bibitem[Ginsberg et~al.(2009)Ginsberg, Mohebbi, Patel, Brammer, Smolinski, and
  Brilliant]{ginsberg2009detecting}
Jeremy Ginsberg, Matthew~H. Mohebbi, Rajan~S. Patel, Lynnette Brammer, Mark~S.
  Smolinski, and Larry Brilliant.
\newblock Detecting influenza epidemics using search engine query data.
\newblock \emph{Nature}, 457\penalty0 (7232):\penalty0 1012--1014, 2009.

\bibitem[Grefenstette et~al.(2013)Grefenstette, Brown, Rosenfeld, DePasse,
  Stone, Cooley, Wheaton, Fyshe, Galloway, Sriram, Guclu, Abraham, and
  Burke]{grefenstette2013fred}
John~J. Grefenstette, Shawn~T. Brown, Roni Rosenfeld, Jay DePasse, Nathan~T.B.
  Stone, Phillip~C. Cooley, William~D. Wheaton, Alona Fyshe, David~D. Galloway,
  Anuroop Sriram, Hasan Guclu, Thomas Abraham, and Donald~S. Burke.
\newblock {FRED} (a {F}ramework for {R}econstructing {E}pidemic {D}ynamics): an
  open-source software system for modeling infectious diseases and control
  strategies using census-based populations.
\newblock \emph{BMC public health}, 13\penalty0 (1):\penalty0 940, 2013.

\bibitem[{Health and Human Services}(2015)]{hhs}
{Health and Human Services}.
\newblock Regional offices, 2015.
\newblock URL
  \url{https://www.hhs.gov/about/agencies/regional-offices/index.html?language=es}.
\newblock Accessed: 03-29-2017.

\bibitem[Hickmann et~al.(2015)Hickmann, Fairchild, Priedhorsky, Generous,
  Hyman, Deshpande, and Del~Valle]{hickmann2015forecasting}
Kyle~S. Hickmann, Geoffrey Fairchild, Reid Priedhorsky, Nicholas Generous,
  James~M. Hyman, Alina Deshpande, and Sara~Y. Del~Valle.
\newblock Forecasting the 2013--2014 influenza season using {W}ikipedia.
\newblock \emph{PLoS Computational Biology}, 11\penalty0 (5):\penalty0
  e1004239, 2015.

\bibitem[Higdon et~al.(2008)Higdon, Gattiker, Williams, and
  Rightley]{higdon2008computer}
Dave Higdon, James Gattiker, Brian Williams, and Maria Rightley.
\newblock Computer model calibration using high-dimensional output.
\newblock \emph{Journal of the American Statistical Association}, 103\penalty0
  (482):\penalty0 570--583, 2008.

\bibitem[Huang et~al.(2014)Huang, Lipsitch, Shaman, and Goldstein]{huang2014us}
Karen~E. Huang, Marc Lipsitch, Jeffrey Shaman, and Edward Goldstein.
\newblock The {US} 2009 {A}/{H}1{N}1 influenza epidemic: quantifying the impact
  of school openings on the reproductive number.
\newblock \emph{Epidemiology (Cambridge, Mass.)}, 25\penalty0 (2):\penalty0
  203, 2014.

\bibitem[Kennedy and {O'Hagan}(2001)]{kennedy2001bayesian}
Marc~C. Kennedy and Anthony {O'Hagan}.
\newblock Bayesian calibration of computer models.
\newblock \emph{Journal of the Royal Statistical Society: Series B (Statistical
  Methodology)}, 63\penalty0 (3):\penalty0 425--464, 2001.

\bibitem[Kermack and McKendrick(1927)]{kermack1927contribution}
William~O. Kermack and Anderson~G. McKendrick.
\newblock A contribution to the mathematical theory of epidemics.
\newblock In \emph{Proceedings of the Royal Society of London A: mathematical,
  physical and engineering sciences}, pages 700--721. The Royal Society, 1927.

\bibitem[Lazer et~al.(2014)Lazer, Kennedy, King, and
  Vespignani]{lazer2014parable}
David Lazer, Ryan Kennedy, Gary King, and Alessandro Vespignani.
\newblock The parable of {G}oogle flu: traps in big data analysis.
\newblock \emph{Science}, 343\penalty0 (6176):\penalty0 1203--1205, 2014.

\bibitem[Linzer(2013)]{linzer2013}
Drew~A. Linzer.
\newblock Dynamic {B}ayesian forecasting of presidential elections in the
  {S}tates.
\newblock \emph{Journal of the American Statistical Association}, 108\penalty0
  (501):\penalty0 124--134, 2013.

\bibitem[Mniszewski et~al.(2008)Mniszewski, Del~Valle, Stroud, Riese, and
  Sydoriak]{mniszewski2008episims}
Susan~M. Mniszewski, Sara~Y. Del~Valle, Phillip~D. Stroud, Jane~M. Riese, and
  Stephen~J. Sydoriak.
\newblock Episims simulation of a multi-component strategy for pandemic
  influenza.
\newblock In \emph{Proceedings of the 2008 Spring simulation multiconference},
  pages 556--563. Society for Computer Simulation International, 2008.

\bibitem[Molinari et~al.(2007)Molinari, Ortega-Sanchez, Messonnier, Thompson,
  Wortley, Weintraub, and Bridges]{molinari2007annual}
Noelle-Angelique~M. Molinari, Ismael~R. Ortega-Sanchez, Mark~L. Messonnier,
  William~W. Thompson, Pascale~M. Wortley, Eric Weintraub, and Carolyn~B.
  Bridges.
\newblock The annual impact of seasonal influenza in the {US}: measuring
  disease burden and costs.
\newblock \emph{Vaccine}, 25\penalty0 (27):\penalty0 5086--5096, 2007.

\bibitem[Nsoesie et~al.(2014)Nsoesie, Brownstein, Ramakrishnan, and
  Marathe]{nsoesie2014systematic}
Elaine~O. Nsoesie, John~S. Brownstein, Naren Ramakrishnan, and Madhav~V.
  Marathe.
\newblock A systematic review of studies on forecasting the dynamics of
  influenza outbreaks.
\newblock \emph{Influenza and other respiratory viruses}, 8\penalty0
  (3):\penalty0 309--316, 2014.

\bibitem[Osthus et~al.(2017)Osthus, Hickmann, Caragea, Higdon, and
  Del~Valle]{osthus2017flu}
Dave Osthus, Kyle~S. Hickmann, Petru\c{t}a~C. Caragea, Dave Higdon, and Sara~Y.
  Del~Valle.
\newblock Forecasting seasonal influenza with a state-space {SIR} model.
\newblock \emph{Annals of Applied Statistics}, 11\penalty0 (1):\penalty0
  202--224, 2017.

\bibitem[Plummer(2003)]{plummer2003}
Martyn Plummer.
\newblock {JAGS}: A program for analysis of {B}ayesian graphical models using
  {G}ibbs sampling.
\newblock In \emph{Proceedings of the 3rd international workshop on distributed
  statistical computing}, volume 124, page 125. Vienna, 2003.

\bibitem[Plummer(2016)]{rjags}
Martyn Plummer.
\newblock \emph{rjags: {B}ayesian Graphical Models using MCMC}, 2016.
\newblock URL \url{https://CRAN.R-project.org/package=rjags}.
\newblock R package version 4-6.

\bibitem[Plummer et~al.(2006)Plummer, Best, Cowles, and Vines]{coda}
Martyn Plummer, Nicky Best, Kate Cowles, and Karen Vines.
\newblock {CODA}: Convergence diagnosis and output analysis for {MCMC}.
\newblock \emph{R News}, 6\penalty0 (1):\penalty0 7--11, 2006.
\newblock URL \url{http://CRAN.R-project.org/doc/Rnews/}.

\bibitem[Polgreen et~al.(2008)Polgreen, Chen, Pennock, Nelson, and
  Weinstein]{polgreen2008using}
Philip~M. Polgreen, Yiling Chen, David~M. Pennock, Forrest~D. Nelson, and
  Robert~A. Weinstein.
\newblock Using {I}nternet searches for influenza surveillance.
\newblock \emph{Clinical infectious diseases}, 47\penalty0 (11):\penalty0
  1443--1448, 2008.

\bibitem[Priedhorsky et~al.(2017)Priedhorsky, Osthus, Daughton, Moran,
  Generous, Fairchild, Deshpande, and Del~Valle]{priedhorsky2017measuring}
Reid Priedhorsky, Dave Osthus, Ashlynn~R. Daughton, Kelly~R. Moran, Nicholas
  Generous, Geoffrey Fairchild, Alina Deshpande, and Sara~Y. Del~Valle.
\newblock Measuring global disease with {W}ikipedia: Success, failure, and a
  research agenda.
\newblock In \emph{Proceedings of the 2017 ACM Conference on Computer Supported
  Cooperative Work and Social Computing}, pages 1812--1834. ACM, 2017.

\bibitem[{R Core Team}(2016)]{R}
{R Core Team}.
\newblock \emph{R: A Language and Environment for Statistical Computing}.
\newblock R Foundation for Statistical Computing, Vienna, Austria, 2016.
\newblock URL \url{https://www.R-project.org/}.

\bibitem[Rolfes et~al.(2016)Rolfes, Foppa, Garg, Flannery, Brammer, Singleton,
  Burns, Jernigan, Reed, Olsen, and Bresee]{fluburden2016}
Melissa~A. Rolfes, Ivo~M. Foppa, Shikha Garg, Brendan Flannery, Lynnette
  Brammer, James~A. Singleton, Erin Burns, Daniel Jernigan, Carrie Reed,
  Sonja~J. Olsen, and Joseph Bresee.
\newblock Estimated influenza illnesses, medical visits, hospitalizations, and
  deaths averted by vaccination in the {U}nited {S}tates, 2016.
\newblock URL \url{https://www.cdc.gov/flu/about/disease/2015-16.htm}.
\newblock Accessed: 04-01-2017.

\bibitem[Shaman and Karspeck(2012)]{shaman2012}
Jeffrey Shaman and Alicia Karspeck.
\newblock Forecasting seasonal outbreaks of influenza.
\newblock \emph{Proceedings of the National Academy of Sciences}, 109\penalty0
  (50):\penalty0 20425--20430, 2012.

\bibitem[Shaman et~al.(2013)Shaman, Karspeck, Yang, Tamerius, and
  Lipsitch]{shaman2013}
Jeffrey Shaman, Alicia Karspeck, Wan Yang, James Tamerius, and Marc Lipsitch.
\newblock Real-time influenza forecasts during the 2012--2013 season.
\newblock \emph{Nature communications}, 4, 2013.

\bibitem[Soebiyanto et~al.(2010)Soebiyanto, Adimi, and
  Kiang]{soebiyanto2010modeling}
Radina~P. Soebiyanto, Farida Adimi, and Richard~K. Kiang.
\newblock Modeling and predicting seasonal influenza transmission in warm
  regions using climatological parameters.
\newblock \emph{PloS one}, 5\penalty0 (3):\penalty0 e9450, 2010.

\bibitem[Tetlock et~al.(2017)Tetlock, Mellers, and
  Scoblic]{tetlock2017bringing}
Philip~E. Tetlock, Barbara~A. Mellers, and J.~Peter Scoblic.
\newblock Bringing probability judgments into policy debates via forecasting
  tournaments.
\newblock \emph{Science}, 355\penalty0 (6324):\penalty0 481--483, 2017.

\bibitem[Towers and Feng(2009)]{towers2009pandemic}
Sherry Towers and Zhilan Feng.
\newblock Pandemic {H}1{N}1 influenza: Predicting the course of vaccination
  programme in the {U}nited {S}tates.
\newblock \emph{Euro. Surveill}, 14, 2009.

\bibitem[Viboud et~al.(2003)Viboud, Bo{\"e}lle, Carrat, Valleron, and
  Flahault]{viboud2003prediction}
C{\'e}cile Viboud, Pierre-Yves Bo{\"e}lle, Fabrice Carrat, Alain-Jacques
  Valleron, and Antoine Flahault.
\newblock Prediction of the spread of influenza epidemics by the method of
  analogues.
\newblock \emph{American Journal of Epidemiology}, 158\penalty0 (10):\penalty0
  996--1006, 2003.

\bibitem[Weiss(2013)]{weiss2013sir}
Howard~H. Weiss.
\newblock The {SIR} model and the foundations of public health.
\newblock \emph{Materials matem{\`a}tics}, pages 1--17, 2013.

\bibitem[Wilson and Brownstein(2009)]{wilson2009early}
Kumanan Wilson and John~S. Brownstein.
\newblock Early detection of disease outbreaks using the {I}nternet.
\newblock \emph{Canadian Medical Association Journal}, 180\penalty0
  (8):\penalty0 829--831, 2009.

\end{thebibliography}

\end{document}